\documentclass{article}
\usepackage{amssymb,amsmath,amscd,amsfonts,euscript}
\usepackage{amsthm}
\usepackage{fancyhdr}
\usepackage{xcolor}
\usepackage[colorlinks=true, urlcolor=blue, pdfborder={0 0 0}]{hyperref}
\usepackage{subcaption}
\usepackage{fancyhdr}
\usepackage{graphicx}
\usepackage{enumerate}
\usepackage{multicol}
\usepackage{srcltx}
\newtheorem{theorem}{Theorem}[section]
\newtheorem{proposition}{Proposition}[section]
\newtheorem{lemma}{Lemma}[section]

\newtheorem{corollary}{Corollary}[section]

\newtheorem{remark}{Remark}
\newcommand{\wh}{\widehat}

\newcommand{\ov}{\overline}

\newcommand{\wt}{\widetilde}

\newcommand\cE{{\cal E}}

\newcommand\cF{{\cal F}}

\newcommand\cN{{\cal N}}

\newcommand\e{{\varepsilon}}

\newcommand{\bbr}{{\mathbb R}}

\def\endproof{\mbox{\ $\qed$}}
\def\sign{\mbox{sign}}

\def\E{{\bf E}}

\def\P{{\bf P}}
\def\C{{\bf C}}
\def\L{{\bf L}}

\def\A{{\bf D}}

\def\A{{\bf A}}

\def\H{{\bf H}}

\def\Q{{\bf Q}}
\def\r{{\bf r}}

\def\a{{\bf a}}

\def\v{{\bf v}}

\def\Chi{{\bf 1}}
\def\d{\mathrm{d}}
\newcommand{\zs}[1]{{\mathchoice{#1}{#1}{\lower.25ex\hbox{$\scriptstyle#1$}}
{\lower0.25ex\hbox{$\scriptscriptstyle#1$}}}}

\begin{document}

\title{Approximate hedging problem with transaction costs in stochastic volatility markets\thanks{The second author is partially supported by Russian Science Foundation (research project No. 14-49-00079)
and by  the National Research 
Tomsk State University.
}}

\author{
Thai Huu Nguyen\thanks{Laboratoire de Math\'{e}matiques Rapha\"{e}l Salem, 
UMR 6085 CNRS-Universit\'{e} de Rouen, France and
Department of Mathematics and Statistics, University of Economics, Ho Chi Minh City, Vietnam, 
e-mail: thaibopy@gmail.com, huu.nguyen@etu.univ-rouen.fr.} and 
Serguei Pergamenshchikov
\thanks{Laboratoire de Math\'{e}matiques Rapha\"{e}l Salem, UMR 6085 CNRS-Universit\'{e} de Rouen, France, and National Research University - Higher School of Economics, Laboratory of Quantitative Finance,
Moscow, Russia,
 e-mail: serge.pergamenchtchikov@univ-rouen.fr.}
}

\maketitle
\begin{abstract} This paper studies the problem of option replication in general stochastic volatility markets with transaction costs, using a new specification for the volatility adjustment in Leland's algorithm \cite{Leland}. We prove several limit theorems for the normalized replication error of Leland's strategy, as well as that of the strategy suggested by L\'epinette \cite{Lepinette10}. The asymptotic results obtained not only generalize the existing results, but also enable us to fix the under-hedging property pointed out by Kabanov and Safarian in \cite{Kab-Saf1}. We also discuss possible methods to improve the convergence rate and to reduce the option price inclusive of transaction costs.
\end{abstract}


\vspace{3mm}


\section{Introduction}
Leland \cite{Leland} suggests a simple method for pricing standard European options in markets
with proportional transaction costs. He argues that transaction costs can be accounted for in the option price by increasing the volatility parameter in the classical Black-Scholes model \cite{BlackScholes}. Leland then claims, without giving a mathematically rigorous proof, that the replicating portfolio of the corresponding discrete delta strategy converges to the option payoff as the number of revisions $n$ goes to infinity, if the transaction cost rate is a constant independent of $n$, or decreases to zero at the rate $n^{-1/2}$. The latter statement is proved by Lott in his PhD thesis \cite{Lott}. In fact, this property still holds if the transaction cost coefficient converges to zero at any power rate \cite{Kab-Saf1}.

However, a careful analysis shows that the replicating portfolio does not converge to the option payoff when the cost rate is a constant independent of $n$. Kabanov and Safarian \cite{Kab-Saf1} find an explicit limit for the hedging error, which is negative, showing that the replication problem is not completely solved in Leland's framework. Pergamenshchikov \cite{Per} obtains a weak convergence for the normalized hedging error and points out that, for the case of constant transaction cost, the rate of convergence in Kabanov-Safarian's result is $n^{1/4}$. This limit theorem is of practical importance because it provides the asymptotic distribution of the hedging error. Note that the rate of convergence can be improved using non-uniform revisions \cite{Lepinette10, Darses}. In these papers, L\'epinette and his co-authors suggest a modification to Leland's strategy to solve the discrepancy identified by Kabanov and Safarian. For a recent account of the theory, we refer the reader to Section \ref{sec:Rev} and \cite{Kab-Saf1,Lepinette08,Lepinette-Kab,Lepinette10,Gam,Grand,Darses,Per}.

In this study, we examine the problem of approximate hedging of European style options in stochastic volatility (SV) markets with constant transaction costs (the reader is referred to e.g. \cite{Fouque} and the
references therein for motivations and detailed discussions related to SV models). In particular, we establish a weak convergence for the normalized hedging error of
 Leland's strategy using a simple volatility adjustment, in a general SV setting. The results obtained not only generalize the existing results, but also provide a method for improving the rate of convergence. Furthermore, it turns out that superhedging can be attained by controlling a model parameter. 

Let us emphasize that the classic form for {\it adjusted volatility} proposed in \cite{Leland} and applied in \cite{Kab-Saf1,Kab-Saf2,Lepinette08,Lepinette09,Lepinette10} may not be applicable in SV models. The reason is that option pricing and hedging are intrinsically different in SV markets than in the classical Black-Scholes framework. In particular, the option price now depends on future realizations of the volatility process. In general, this information may not be statistically available for all investors. To treat this issue, we suggest a new specification for adjusted volatility in Leland's algorithm. Although we employ an artificially modified volatility, simpler than the well-known version used in the previous literature, the same asymptotic results are obtained for SV contexts. In addition, the rate of convergence can be improved by controlling a model parameter. Note that, in the above-mentioned papers, approximation procedures are mainly based on moment estimates. This essential technique no longer works in general SV models, unless some intrinsic conditions are imposed on the model parameters \cite{And-Pit, Lions-Mus}. It is useful to remember that our goal is to establish a weak convergence for the normalized replicating error which only requires convergence in probability of the approximation terms. Thus, in the approximation procedure, the integrability issue can be avoided in order to keep our model setting as general as possible.

As discussed in \cite{Per}, the option price (inclusive of transaction costs) in Leland's algorithm may be
high (it, in fact, approaches the buy-and-hold price), even for small values of the revision number. Another practical advantage of our method is that the option price can be reduced as long as the option
seller is willing to take a risk in option replication. This approach is
inspired by the theory of quantile hedging \cite{Follmer}.

\vspace{1mm}
The remainder of the paper is organized as follows. In Section \ref{sec:Rev}, we give a brief review of Leland's approach. Section \ref{Mar} is devoted to formulating the problem and presenting our main results. Section 
\ref{sec:Ap} presents some direct applications to pricing and hedging. Section \ref{Exa} discusses common SV models that fulfill our condition on the volatility function. A numerical result for Hull-White's model is also provided for illustration. Section \ref{hfm} connects our results to high-frequency markets with proportional transaction costs. The proofs of our main results are reported in Section \ref{Proofs}. Auxiliary lemmas can be found in the Appendix.

\section{Approximate hedging with transaction costs: A review of Leland's approach}
\label{sec:Rev}

In a complete no-arbitrage model (i.e., there exists a unique equivalent
martingale measure under which the stock price is a martingale), options can
be completely replicated by a self-financing trading strategy. The option price,
defined as the replication cost, is the initial capital that the investor
must invest to obtain a complete hedge. In fact, the option price can be
computed as the expectation of the discounted claim under the unique equivalent
martingale measure. This principle plays a central role in the well-known Black-Scholes model. For simplicity, let us consider a continuous time  model of a two-asset financial market 
on the time interval $[0,1]$, where the bond price is equal to 1 at all times. The stock price dynamics follow the stochastic
differential equation
\begin{equation}\label{sec:Rev.1}
\d S_\zs{t}=\sigma_\zs{0}S_\zs{t}\d W_\zs{t}\,, \quad S_0 \quad \mbox{given},
\end{equation}
where $S_0$ and $\sigma_\zs{0}$ are positive constants and $(W_\zs{t})_\zs{0\le t\le 1}$ is
a standard Wiener process. As usual, let
$\cF_\zs{t}=\sigma\{W_\zs{u}\,,\,0\le u\le t\}$.  We recall that a financial strategy 
$(\beta_\zs{t},\gamma_\zs{t})_\zs{0\le t\le 1} $ is an \emph{admissible
self-financing strategy} if it is bounded from below, $(\cF_\zs{t})$ - adapted with $
\int_\zs{0}^{t}(|\beta_\zs{t}|+\gamma_\zs{t}^{2})
\,\d t<\infty$ a.s., 
and the portfolio value satisfies
$$
V_\zs{t}=\beta_\zs{t}+\gamma_\zs{t}S_\zs{t}=V_\zs{0}+\int_\zs{0}^{t}\gamma_\zs{u}\d S_\zs{u},\,\quad t\in [0,1].
$$
The classic hedging problem is to find an admissible self-financing strategy $( \beta_\zs{t},\gamma_\zs{t})$ whose terminal portfolio value exceeds the payoff $h(S_\zs{1})=\max(S_\zs{1}-K,0)$, or
$$
V_\zs{1}=
V_\zs{0}+\int_\zs{0}^{1}\gamma_\zs{u}\d S_\zs{u}\ge
h(S_\zs{1})
\qquad \mbox{a.s.},
$$
where $K$ is the strike price. The standard pricing principle shows that the option price $C( t,S_\zs{t})$
is given by the well-known formula \cite{BlackScholes}  
\begin{equation}\label{sec:Rev.2-1}
C(t,x)=
C(t,x,\sigma_\zs{0}) =x\Phi (\wt{\v}(t,x)) -K\Phi (
\wt{\v}(t,x)-\sigma_\zs{0}\sqrt{1-t})\,,
\end{equation}
where
\begin{equation}\label{sec:Rev.2}
\wt{\v}(t,x)=\v(\sigma^{2}_\zs{0}(1-t),x)\quad
\mbox{and}\quad
\v(\lambda,x)=\frac{\ln(x/K)}{\sqrt{\lambda}}\,
+
\frac{\sqrt{\lambda}}{2}\,.
\end{equation}
Here,
$\Phi$ is the standard normal distribution function.
In the following, we denote by 
$\varphi$ the $\cN(0,1)$ density: $\varphi(z)=\Phi'(z)$.
 One can check
directly that 
\begin{equation}\label{sec:Rev.3}
C_\zs{x}(t,x)=\Phi(\wt{\v}(t,x))
\quad\mbox{and}\quad
C_\zs{xx}(t,x)=
\frac{\varphi(\wt{\v}(t,x))}{x\sigma_\zs{0}\sqrt{1-t}}
\,.
\end{equation}
By assuming that continuous portfolio adjustments are possible with zero transaction costs, Black and Scholes \cite{BlackScholes} argue that the option payoff can be dynamically replicated using the delta strategy (i.e., the partial derivative of the option price with
respect to the stock price).

It is clear that the assumption of continuous portfolio revision is not realistic. Moreover, continuous trading would be ruinously expensive in the case of nonzero constant proportional transaction costs because the delta strategy has infinite variation. This simple intuition contradicts the argument of Black and Scholes that, if trading takes places reasonably frequently, then hedging errors are relatively small. Therefore, option
pricing and replication with nonzero trading costs are intrinsically different from those in the Black-Scholes setting. Note that it may be very costly to assure a given degree of accuracy in replication with transaction costs. In what follow, we show that Leland's increasing
volatility principle \cite{Leland} is practically helpful in such contexts.

\subsection{Constant volatility case}\label{Const}
Leland's approach \cite{Leland}
provides an efficient technique to deal with transaction costs. This method is simply based on the intuition that
transaction costs can be accounted for in the option price as a reasonable extra fee, necessary for the option seller to cover the option return. It means that in the presence of transaction costs, the option becomes more expensive than
in the classic Black-Scholes framework. This is intuitively equivalent to an increase in
the volatility parameter in the Black-Scholes formula.
%\subsection{Constant volatility case}\label{subsec:const}
Let us shortly describe Leland's approach \cite{Leland,Kab-Saf1}. Suppose that for each trading activity, the investor has to pay a fee directly proportional to the trading volume, measured in dollar value. Assume that the 
 transaction cost rate is given by the 
law $\kappa_\zs{*}n^{-\alpha}$, where $n$ is the number of revisions. Here, $0\le\alpha\le 1/2$ and  $\kappa_\zs{*}>0$ are two fixed parameters. The basic idea of Leland's method is to replace the true volatility parameter in the Black-Scholes model by $\wh{\sigma}$, artificially modified as
\begin{equation}\label{sec:Rev.5}
\wh{\sigma}^{2}=\sigma_\zs{0}^{2}+
\varrho
\,n^{1/2-\alpha}
\quad\mbox{with}\quad
\varrho =\kappa_\zs{*} \sigma_\zs{0}\sqrt{8/\pi }
\,.
\end{equation}
In this case, the option price is given by $\wh{C}
( t,x) =C( t,x,\wh{
\sigma})$, the Black-Scholes's formula. For the problem of option replication, Leland suggests the following discrete strategy, known as Leland's strategy,
\begin{equation}\label{sec:Rev.4-1}
\gamma_\zs{t}^{n}=\sum_\zs{i=1}^{n}\wh{C}_\zs{x}( t_\zs{i-1},S_\zs{t_\zs{i-1}})
{\bf 1}_\zs{(t_\zs{i-1},t_\zs{i}]}( t), \quad t_\zs{i}=\frac{i}{n},\ i\in\{1,2,..,n\}.
\end{equation}
%We assume further that
%the portfolio is revised discretely at 
%%\begin{equation}\label{sec:Rev.4}
%$t_\zs{i}=\frac{i}{n},\ i\in\{1,2,..,n\},$
%%\end{equation}
Here, the number of shares held in the interval $(t_\zs{i-1},t_\zs{i}]$ is the delta strategy calculated at the left bound of this interval.
Then, the replicating portfolio value takes the form
\begin{equation}\label{sec:Rev.6}
V_\zs{1}^{n}=V_\zs{0}^{n}+\int_\zs{0}^{1}\gamma_\zs{u}^{n}\d S_\zs{u}-
\kappa_\zs{*}n^{-\alpha}\,J_\zs{n}\,,
\end{equation}
where the total trading volume is
%\begin{equation}\label{sec:Rev.6-1}
$J_\zs{n}=\sum^{n}_\zs{i=1}\,S_\zs{t_\zs{i}}\vert \gamma_\zs{t_\zs{i}}^{n}-\gamma
_\zs{t_\zs{i-1}}^{n}\vert$
%\end{equation}
(measured in dollar value). Recall that the option price $\wh{C}(t,x) $ is the solution of the Black-Scholes
PDE with the adjusted volatility $\wh{\sigma}$
\begin{equation}\label{sec:Rev.7}
\wh{C}_\zs{t}(t,x)+\dfrac{1}{2}\wh{\sigma}^{2}x^{2}\wh{C}
_\zs{xx}( t,x) =0\,,\quad 0\le t<1; \quad\wh{C}(1,x)=h(x)\,.
\end{equation}
Using It\^o's formula, we can represent the tracking error, $V_\zs{1}^{n}-h(S_\zs{1})$, as
\begin{align}\label{sec:Rev.8}
\int_\zs{0}^{1}\left(\gamma_\zs{t}^{n}-\wh{C}_\zs{x}(t,S_\zs{t})
\right)\d S_\zs{t}%\\[2mm]
+\frac{1}{2}(\wh{\sigma}^{2}-\sigma_{0}^{2})\int_\zs{0}^{1}S_\zs{t}^{2}\wh{C}_\zs{xx}(
t,S_\zs{t})\d t-\kappa 
_\zs{*}n^{-\alpha}J_\zs{n}\,.
\end{align}
\begin{remark}[Leland]\label{Re.sec:Rev.1-0}
The specific form \eqref{sec:Rev.5} results from the following intuition: the
Lebesgue integral in \eqref{sec:Rev.8} is clearly well approximated by
the Riemann sum of the terms $\sigma_\zs{0}S_\zs{t_\zs{i-1}}^{2}\wh{C}_\zs{xx}(t_\zs{i-1},S_\zs{t_\zs{i-1}}){\Delta t} $, while $S_\zs{t_\zs{i}}\vert \gamma_\zs{t_\zs{i}}^{n}-\gamma_\zs{t_\zs{i-1}}^{n}\vert$ can be replaced by
\begin{equation*}
\approx \sigma_\zs{0}S^2_\zs{t_\zs{i-1}}\wh{C}_\zs{xx}(
t_\zs{i-1},S_\zs{t_\zs{i-1}})\vert
\Delta W_\zs{t_\zs{i}}\vert \approx \sigma_\zs{0}
\sqrt{2/(n\pi)}\,
S^2_\zs{t_\zs{i-1}}\,
\wh{C}_\zs{xx}(t_\zs{i-1},S_\zs{t_\zs{i-1}}),
\end{equation*}
because $\E\vert\Delta W_\zs{t_\zs{i}}\vert 
=\sqrt{2/\pi }\sqrt{\Delta t}=\sqrt{2/(\pi n)}$. 
Hence, it is reasonable to expect that the modified volatility defined in \eqref{sec:Rev.5} will give an appropriate approximation to compensate transaction costs.
\end{remark}

Leland \cite{Leland} conjectures that the replication error converges in probability to zero as $ n\to\infty$ for the case of constant proportional transaction cost (i.e., $\alpha=0$).  
He also suggests, without giving a rigorous proof, that this property is also true for the case
$\alpha=1/2$. In fact, Leland's second conjecture for $\alpha=1/2$ is correct and is proved by Lott in
his PhD thesis \cite{Lott}.
%\subsection{Leland-Lott result and extension for the case $\alpha=1/2$}\label{1/2}
\begin{theorem}[Leland-Lott \cite{Leland,Lott}]\label{Th.sec:Rev.1}
For $\alpha =1/2$, strategy \eqref{sec:Rev.4-1} 
defines an approximately
replicating portfolio as the number of revision intervals $n$ tends
to infinity
$$
\P-\lim_\zs{n\rightarrow \infty }V_\zs{1}^{n}=
h(S_\zs{1})\,.
$$
\end{theorem}

\noindent This result is then extended by Ahn \emph{et al.} in \cite{Ahn} to general diffusion models. Kabanov and Safarian \cite{Kab-Saf1} observe that the Leland-Lott theorem remains true 
as long as the cost rate converges to zero as 
$n\rightarrow \infty$.

\begin{theorem}[Kabanov-Safarian \cite{Kab-Saf1}]\label{Th.sec:Rev.2}
For any $0<\alpha\le 1/2$,
$
\P-\lim_\zs{n\rightarrow \infty }V_\zs{1}^{n}=
h(S_\zs{1})\,.
$
\end{theorem}
In \cite{Lepinette-Kab,Kab-Saf2}, the authors study the Leland-Lott approximation in the sense
of $L^{2}$ convergence for the case $\alpha =1/2$.\footnote{ Seemingly, mean-square replication may not contain much useful information because gains and losses have different meaning in practice. Clearly, if $\alpha =1/2$ the modified
volatility is independent of $n$.}
\begin{theorem}[Kabanov-L\'epinette \cite{Lepinette-Kab}]\label{Th.sec:Rev.5}
Let $\alpha =1/2$. The mean-square approximation error of Leland's strategy, with $\varrho$ defined in
\eqref{sec:Rev.5},
satisfies the following asymptotic equality
$$
\E\left( V_\zs{1}^{n}-h(S_\zs{1}) \right)^{2}=An^{-1}+o(
n^{-1})\quad \mbox{as}\quad n\rightarrow \infty,
$$
where $A$ is some positive function.
%$$
%B=\int_\zs{0}^{1}\left( \frac{\sigma_\zs{0}^{4}}{2}+\kappa_\zs{*}\sigma_\zs{1}^{3}
%\sqrt{\frac{2}{\pi }}+\kappa_\zs{*}^{2}\sigma_\zs{0}^{2}( 1-\frac{2}{\pi }
%) \right) \Lambda_\zs{t}dt,
%$$
%with $\Lambda_\zs{t}$ is explicitly determined by the
%expectation of $S_\zs{t}^{4}\wh{C}_\zs{xx}( t,S_\zs{t}) $.
\end{theorem}
Theorem \ref{Th.sec:Rev.5} suggests that the normalized replication error converges in law as $n\to\infty$.\begin{theorem}[L\'epinette-Kabanov \cite{Kab-Saf2}]\label{Th.sec:Rev.5-1}
For $\alpha=1/2$, the processes $Y^n=n^{1/2}( V^{n}-h(S_\zs{1}))$ converge weakly in the Skorokhod space ${\cal D}[0,1]$ to the distribution of the process $Y_{\bullet}=\int_\zs{0}^{\bullet} B(S_\zs{t}) \d Z_\zs{t}$, where $Z$ is an independent Wiener process.
\end{theorem}
\begin{remark}
An interesting connection between this case and the problem of hedging under proportional transaction costs in high-frequency markets is discussed in Section \ref{hfm}. 
\end{remark}
%\subsection{The case of constant proportional cost $\alpha=0$}
It is important to note that Leland's approximation 
in Remark~\ref{Re.sec:Rev.1-0} is not mathematically correct and thus, 
his first conjecture is not valid for the case of constant transaction costs. 
In fact, as $n\to\infty$, the trading volume
$J_\zs{n}$ can be approximated by the following sum (which converges in probability to $J(S_\zs{1},\varrho)$ defined in \eqref{sec:Rev.4-3})
%\begin{equation}\label{sec:Rev.4-2-0}
$$-\sum_{i=1}^{n}
{\lambda^{-1/2}_\zs{i-1}}S_\zs{t_\zs{i-1}}\,
\wt{\varphi}(\lambda_\zs{i-1},S_\zs{t_\zs{i-1}})
| 
{\sigma_\zs{0}}{\varrho}^{-1}
Z_\zs{i}+q(\lambda_\zs{i-1},S_\zs{t_\zs{i-1}})
|
\,\Delta \lambda_\zs{i}\,,$$
%\end{equation} 
where 
$\lambda_\zs{i}=\lambda_\zs{t_{i}}=\wh{\sigma}^{2}(1-t_\zs{i}),$ $
Z_\zs{i}=\Delta W_\zs{t_\zs{i}}/\sqrt{\Delta t_\zs{i}}
$  and 
\begin{equation}\label{sec:Rev.4-2-1}
\wt{\varphi}(\lambda,x) = \varphi(\v(\lambda,x)),\quad q(\lambda,x)=\frac{\ln(x/K)}{2\lambda}-\frac{1}{4}\,.
\end{equation}
A careful study confirms that there is a \emph{non trivial discrepancy} between 
the limit of the replicating portfolio and the payoff for the case $\alpha=0$.

\begin{theorem}[Kabanov-Safarian \cite{Kab-Saf1}]\label{Th.sec:Rev.3}
For $\alpha=0$, $
V_\zs{1}^{n}$ converges to $h(S_\zs{1})+
\min (S_\zs{1},K) -\kappa_\zs{*}J(S_\zs{1},\varrho)\,
$ in probability, where
\begin{equation}\label{sec:Rev.4-3}
J(x,\varrho) =x\int_\zs{0}^{+\infty }
\,
\lambda ^{-1/2}\wt{\varphi}(\lambda,x ) 
\,\E\,
\left|
\wt{\varrho}
{Z}
+q(\lambda,x)
\right| 
\,
\d\lambda
\,,
\end{equation}
with $\wt{\varrho}=\sigma_\zs{0}{\varrho}^{-1}$ and ${Z}\sim\cN(0,1)$ independent of $S_1$.
\end{theorem}
\begin{center}
\begin{figure}[h]
\includegraphics[width=1.05\columnwidth,height=7.5cm]{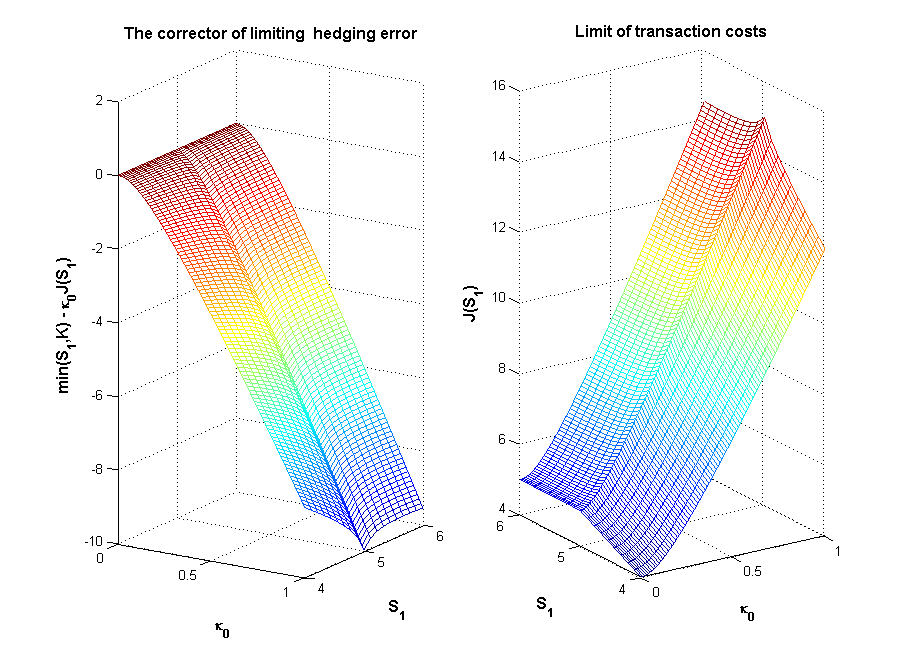}\vspace{-4mm}
\caption{$\min(S_1,K)-\kappa_{*} J(S_1)$ on the left and $J(S_1)$ on the right with $K=5$. }%
\label{Fig.Ch_1.Kab1}%
\end{figure}\end{center}\vspace{-5mm}

\noindent{\bf Under-hedging}: It is important to observe that the problem of option replication is not completely solved in the case of constant transaction costs. Indeed,
considering that $\E\left\vert\wt{\varrho}{Z}\right\vert=1/(2\kappa_\zs{*})$ and the identity
\begin{equation}\label{sec:Rev.4-4}
 x\int\limits_\zs{0}^{\infty }{
\lambda }^{-1/2}\,\wt{\varphi}(\lambda,x) \d\lambda =2\min\,( x,K)
\,,
\end{equation}
we obtain (for the parameter $\varrho$ given in \eqref{sec:Rev.5}) that $\min(x,K) -\kappa_\zs{*}J(x,\varrho)
=x\kappa_\zs{*}$ equals $
\int_\zs{0}^{+\infty }{\lambda}^{-1/2}\wt{\varphi}(\lambda,x)
\,
\left(\E\left\vert \wt{\varrho}{Z}\right\vert-
\E\left\vert 
\wt{\varrho}
{Z}
+q(\lambda,x)\right\vert \right) \d\lambda.
$
Now, Anderson's inequality (see, for example \cite{IbHa}, page 155) implies directly that  
for any  $q\in\bbr$,
$
\E\left\vert \wt{\varrho}{Z}
+q\right\vert \ge\E\left\vert \wt{\varrho} {Z}\right\vert.
$
Therefore, $
\P-
\lim_\zs{n\rightarrow \infty }\,
(V_\zs{1}^{n}-h(S_\zs{1}))\,\le 0,
$
thus, the option is asymptotically under-hedged in this case.

In approximation procedures, one should also pay attention to the fact that $\wh{C}$ and its derivatives depend on the number of revisions when $0\le \alpha<1/2$. In addition, the coefficient 
$\varrho $ appearing in \eqref{sec:Rev.5} can be
chosen in an arbitrary way. 
%The reason of making this parameter varying
%rather than the form \eqref{sec:Rev.5} lies in the fact that the adjusted
%volatility is, indeed, approximate to $n^{1/2}$.
Pergamenshchikov \cite{Per} shows that the rate of convergence in Kabanov-Safarian's theorem is $n^{1/4}$ and provides a weak convergence for the normalized replication error.
\begin{theorem}[Pergamenshchikov \cite{Per}]\label{Th.sec:Rev.4}
Consider Leland's strategy \eqref{sec:Rev.4-1} with $\alpha=0$, and let $\varrho$ in \eqref{sec:Rev.5} be some fixed positive constant. Then, the sequence of random variables
\begin{equation}\label{sec:Rev.4-5}
n^{1/4}( V_\zs{1}^{n}-h(S_\zs{1})
-\min (S_\zs{1},K)
+\kappa_\zs{*}J(S_\zs{1},\varrho) )
\end{equation}
weakly 
converges to a centered mixed Gaussian variable as $n\to\infty$. 
\end{theorem}
%\begin{figure}
%\includegraphics[height=8.5cm,width=13cm]{limit.eps}
%\caption{\small Behavior of limit transaction costs and hedging error.}
%\end{figure}
Theorem \ref{Th.sec:Rev.4} is of practical importance because it not only gives the asymptotic information 
about the hedging error, but also provides a reasonable way to fix the under-hedging 
issue (see Section \ref{sec:Ap}).
%\begin{remark}\label{Re.sec:Rev.2}
%The same result is still valid if revisions are not uniform as disscussed below. 
%Nevertheless, a such result seems to contain little expected information
%because losses and gains clearly have different meanings in practice.
%\end{remark}
Darses and L\'epinette \cite{Darses} modify Leland's strategy in order to improve the convergence rate in Theorem \ref{Th.sec:Rev.4} by applying a non-uniform revision policy $(t_\zs{i})_\zs{1\le i\le n}$, defined by
\begin{equation}\label{sec:Rev.9}
t_\zs{i}=g\left( {i}/{n}\right)\,,\quad 
g(t)=1-(1-t) ^{\mu}
\quad\mbox{for some}\quad \mu\ge 1.
\end{equation}
The adjusted volatility is then taken as 
%\begin{equation}\label{sec:Rev.10}
$\wh{\sigma}^{2}_\zs{t}=\sigma_\zs{0}^{2}+\kappa_\zs{*} \sigma_\zs{0}\sqrt{8/\pi}
\sqrt{n f^{\prime}(t)},$
%\end{equation}
where $f$ is the inverse function of $g$. Furthermore, the discrepancy in Theorems \ref{Th.sec:Rev.3} and \ref{Th.sec:Rev.4} can be removed by employing the following modified strategy, known as L\'epinette's strategy \cite{Lepinette10},
\begin{equation}\label{sec:Rev.9-2}
\bar{\gamma}_\zs{t}^{n}=\sum_\zs{i=1}^{n}
\left(\wh{C}_\zs{x}(t_\zs{i-1},S_\zs{t_\zs{i-1}}) -
\int_\zs{0}^{t_\zs{i-1}}\wh{C}_\zs{xt}(u,S_\zs{u})\d u
\right) 
\Chi_\zs{(t_\zs{i-1},t_\zs{i}]}(t)\,.
\end{equation}

\begin{theorem}\label{Th.sec:Rev.6}
Let $V_\zs{1}^{n}$ be the terminal value of the strategy \eqref{sec:Rev.9-2}
with $\alpha=0$. Then, for any $1\le\mu<\mu_\zs{\max}$, the sequence
$
n^{\beta}( V_\zs{1}^{n}-h(
S_\zs{1}) )
$
weakly converges to a centered mixed Gaussian variable as $n\to\infty$, where
\begin{equation}\label{sec:Rev.9-1}
\beta=\frac{\mu}{2(\mu+1)}
\quad\mbox{and}\quad
\mu_\zs{\max}=\frac{3+\sqrt{57}}{8}
\,.
\end{equation} 
\end{theorem}
 \subsection{Time-dependent volatility case}
Assume now that $\sigma$ is a positive non-random function and the payoff $H$ is a continuous function with continuous derivatives, except at a finite number of
points.  Under the non-uniform rebalancing plan \eqref{sec:Rev.9}, the enlarged volatility should take the form
\begin{equation}\label{sec:Rev.12}
\wh{\sigma}_\zs{t}^{2}=\sigma^{2}(t)+\kappa_\zs{*} 
\sigma(t)n^{1/2-\alpha}\sqrt{f^{\prime}(t) 8/\pi}.
\end{equation}
\begin{theorem}[L\'epinette \cite{Lepinette08}]\label{Th.sec:Rev.7}
Let $\sigma$ be a strictly positive Lipschitz and bounded function. Moreover, suppose that $H(\cdot)$ is a piecewise twice differentiable function and  
there exist $x_\zs{*}\ge 0$ 
and $\delta\ge 3/2$,
such that 
$\sup_\zs{x\ge x_\zs{*}} x^{\delta}\vert H^{''}(x)\vert<\infty$. 
Then, for any $1/2\ge\alpha>0$, the replicating portfolio of Leland's strategy converges in probability to the payoff $H(S_\zs{1})$ as $n\to\infty $. Moreover,
for $\alpha =0$,
$$
\P-\lim_\zs{n\rightarrow \infty }V_\zs{1}^{n}=H(S_\zs{1})
+H_\zs{1}(S_\zs{1}) -\kappa_\zs{*}H_\zs{2}(S_\zs{1}) ,
$$
where $H_\zs{1}(\cdot)$ and $H_\zs{2}(\cdot)$ are positive functions that depend
on the payoff $H$.
\end{theorem}
\begin{remark}\label{Re.sec:Rev.3}
Theorem \ref{Th.sec:Rev.6} still holds in the context of Theorem \ref{Th.sec:Rev.7} (see \cite{Lepinette10}).
\end{remark}

\subsection{Discussion}\label{adjform}

%\subsection{Forms of adjusted volatility for the case $\alpha=0$}

From Remark \ref{Re.sec:Rev.1-0}, the modified volatility defined by \eqref{sec:Rev.5} would seem to give an appropriate approximation that accounts for transaction costs. However, this is not always the case because the option price inclusive of transaction costs now depends on the rebalancing number. In more general models, this specific choice may generate technical issues. For example, in local volatility models \cite{Lepinette08}, proving the existence of the solution to \eqref{sec:Rev.7} requires patience and effort, because $\wh{\sigma}$ depends on the stock price. 
On the other hand, it is interesting to observe that the true volatility plays
 no role in the approximation procedure from a mathematical point of view. In fact, all the results for the case $\alpha=0$ can be obtained by using the form 
$\wh{\sigma}_\zs{t}^{2}=\kappa_\zs{*} 
\sigma(t)n^{1/2}\sqrt{f^{\prime}(t) 8/\pi}$, where the first term $\sigma^{2}(t)$ has been removed. More generally, we can take the following form 
\begin{equation}\label{sec:Rev.14-0}
\wh{\sigma}_\zs{t}^{2}=\varrho \sqrt{nf^{\prime}(t)},
 \end{equation}
for some positive constant $\varrho$, which will be specified later. Of course, the limiting value of transaction costs will change slightly.
%since $\varrho$ is no longer related to the terminal value of volatility, see Theorem \ref{Th.sec:Rev.4}. This important observation follows from the fact which can be proved similarly as Lemma 1.2.8 in \cite{Kab-Saf2} (page 16)
%
%\begin{equation} \label{sec:Rev.15}
%\int_\zs{0}^1\sigma^2(t)S^k_\zs{t}\frac{\partial^k\wh{C}}{\partial x^k}(t,S_\zs{t}) \d t=O(\wh{\sigma}^{-1})=O(n^{-1/4})\quad \mbox{as} \,\, n\to\infty,
%\end{equation}
%for all $k\ge2$. The asymptotic representation \eqref{sec:Rev.15} still holds if $\sigma=\sigma(y_\zs{t})$ for some extra random process $y_\zs{t}$.
Let us emphasize that using the simple form \eqref{sec:Rev.14-0} is important for two reasons. First, it allows us to carry out a far simpler approximation than is used in the existing literature. Second, Leland's strategy with $\wh{\sigma}$ defined in \eqref{sec:Rev.5} may no longer work in stochastic volatility (SV) markets. Indeed, in those markets, option prices depend on future volatility realizations, which are not statistically available. We show in the remainder of the paper, that the simple form \eqref{sec:Rev.14-0} (a deterministic function of $t$) is helpful for approximate hedging in a very general SV setting. It should be noted that the approximation methodology developed here still works well for the classical form \eqref{sec:Rev.5}, if the volatility risk premium depends only on the current value of the volatility process \cite{PhamTouzi,RenaultTouzi}.

We conclude the section by mentioning that Leland's algorithm is of practical importance due to its ease of implementation. The case of constant transaction costs $\alpha =0$ should be investigated in more general situations, for instance, where volatility
depends on external random factors, or jumps in stock
prices are considered. 
%Note that the method of modern approximate hedging theory requires a delicate treatment, which is seemingly impossible in general SV models. 

\section{Model and main results}\label{Mar}

Let $(\Omega ,\mathcal{F}_\zs{1},(\mathcal{F}_\zs{t})_\zs{0\leq
t\leq 1},\P)$ be a standard filtered probability space with
two standard independent $( \mathcal{F}_\zs{t})_\zs{0\leq t\leq 1}$
adapted Wiener processes $(W_\zs{t}^{(1)})$ and 
$(W_\zs{t}^{(2)})$, taking their values in $\mathbb{R}$. Our
financial market consists of one risky asset governed by the following
equations on the time interval $\left[0\,,1\right] $
\begin{equation}\label{sec:Mar.1}
% \left\{ 
% \begin{array}{l}
\d S_\zs{t}=\sigma ( y_\zs{t}) S_\zs{t}\d W_\zs{t}^{(1)}; \quad 
\d y_\zs{t}=F_\zs{1}( t,y_\zs{t}) \d t+F_\zs{2}( t,y_\zs{t})( \r \d W_\zs{t}^{(1)}+\sqrt{1-\r^{2}}\d W_\zs{t}^{(2)}),
% \end{array}
% \right.
\end{equation}
where $-1\le \r \le 1$ is the correlation coefficient. It is well known in the literature of SDEs that if $F_1(t,y)$ and $F_2(t,y)$ are measurable in $(t,y)\in [0,T]\times \bbr$, linearly bounded and locally Lipschitz, there exists a unique solution $y$ to the last equation of system \eqref{sec:Mar.1}. For this fundamental result, see Theorem \ref{Th.sec:Exa.1} and \cite{Fried, Lip-Shir}. For simplicity, assume that the interest rate equals zero. Thus	, the
non-risky asset is chosen as the \emph{num\'{e}raire}. 

In this section, we consider the problem of approximate hedging with constant proportional costs using the principle of increasing volatility for model \eqref{sec:Mar.1}. As discussed in Subsection \ref{adjform}, the adjusted volatility is chosen as
\begin{equation}\label{sec:Mar.2}
\wh{\sigma}_\zs{t}^{2}=\varrho\sqrt{n f^{'}(t)}={\mu}^{-1/2}\varrho \sqrt{n}(1-t)^{\frac{1-\mu }{2\mu}},\quad 1\leq \mu <2.
\end{equation}
The replicating portfolio is revised at $(t_i)$, as defined by \eqref{sec:Rev.9}. The parameter $\varrho >0$ plays an important role in controlling the rate of convergence and is specified later. As shown below, the limiting value of the total trading volume is essentially related to the dependence of $ \varrho $ on the number of revisions. 
%\begin{equation}\label{sec:Mar.2-0}
%\mu_\zs{*}=\frac{1}{2( \mu +2) }\,.
%\end{equation}
\begin{remark}
Intuitively, using an independent adjusted volatility seems unnatural because it fails to account for market information. However, the techniques developed in this note are well adapted to the case where the adjusted volatility depends on a volatility process driven by an independent Brownian motion. In such a context, if the volatility risk premium depends only on the current volatility process, then the no-arbitrage option price (without transaction costs) is the average of the Black-Scholes prices over the future paths of the volatility process \cite{PhamTouzi, RenaultTouzi}.
\end{remark}
Recall that $\wh{C}(t,x)$ is the solution of the Cauchy problem \eqref{sec:Rev.7} with two first derivatives, as described in \eqref{sec:Rev.3}:
$
\wh{C}_\zs{x}(t,x)={\Phi}(\v(\lambda_\zs{t},x))$ and $
\wh{C}_\zs{xx}(t,x)=x^{-1}{\lambda_\zs{t}}^{-1/2}\,
\wt{\varphi}\left(\lambda_\zs{t},x\right),
$
where
\begin{equation}
\lambda_\zs{t}=\int^{1}_\zs{t}\,\wh{\sigma}^{2}_\zs{s}\,\d s
=\wt{\mu}\,\varrho \sqrt{n}(1-t)^{
\frac{1}{4\beta}}
\quad\mbox{and}\quad
 \wt{\mu}=
2\sqrt{\mu}/(\mu+1)\,.
\end{equation}
 \begin{remark}
Section \ref{sec:Ap} will show that the under-hedging situation pointed out in 
\cite{Kab-Saf1} can be fixed by
controlling the parameter $\varrho$.
\end{remark}

We make use of the following condition on the volatility function.\medskip 

\noindent $(\C_\zs{1})$
{\em Assume that $\sigma$ 
is a $C^2$ function and there exists $\sigma_\zs{\min}$ such that
$$
0<\sigma_\zs{\min}\le \sigma(y)\ \mbox{for all}\ y\in \bbr \quad \mbox{and}
\quad \sup_\zs{0\le t\le 1} \E[\sigma^2(y_\zs{t})+\vert \sigma'(y_t)\vert]
<\infty.
$$
}
\noindent Assumption $(\C_1)$ is not restrictive and is fulfilled in many popular SV models (see Section \ref{Exa} and \cite{Pham}).

\subsection{Asymptotic results for Leland's strategy}
Let us study the replication error for Leland's strategy $\gamma_\zs{t}^{n}$ defined in \eqref{sec:Rev.4-1}. 
% \begin{equation*}\label{sec:Mar.3}
% \gamma_\zs{t}^{n}=\sum_\zs{i=1}^{n}\,
% \wt{\Phi}(\lambda_\zs{t_\zs{i-1}}\,, S_\zs{t_\zs{i-1}})\,{\bf 1}_\zs{(t_\zs{i-1},t_\zs{i}]}(
% t).
% \end{equation*}
The replicating portfolio $V^n_1$ is defined by \eqref{sec:Rev.6}.
Now, by It\^{o}'s formula,
\begin{equation}\label{sec:Mar.3-1}
h(S_\zs{1}) =\wh{C}( 1,S_\zs{1})=\wh{C}(
0,S_\zs{0}) +\int_\zs{0}^{1}\wh{C}_\zs{x}( t,S_\zs{t})
\d S_\zs{t}
-\frac{1}{2}I_\zs{1,n}\,,
\end{equation}
where $I_\zs{1,n}=\int_\zs{0}^{1}\,
\left(\wh{\sigma}_\zs{t}^{2}-\sigma ^{2}(
y_\zs{t})\right)\, S_\zs{t}^{2}\wh{C}_\zs{xx}( t,S_\zs{t}) \d t$.
Setting $V_\zs{0}=\wh{C}( 0,S_\zs{0})$, we can represent the replication error
as
\begin{equation}\label{sec:Mar.4-1-0}
V^{n}_\zs{1}-h(S_\zs{1})=\frac{1}{2}I_\zs{1,n}+I_\zs{2,n} -\kappa_\zs{*}J_\zs{n}\,,
\end{equation}
where $I_\zs{2,n}=
\int_\zs{0}^{1}\left(\gamma_\zs{t}^{n}-\wh{C}_\zs{x}(t,S_\zs{t})\right) \d S_\zs{t}$ and 
$J_\zs{n}$ is defined as in \eqref{sec:Rev.6}.
%\vspace{2mm}

Let us first emphasize that complete replication in SV models is far from obvious. In our setting, $I_\zs{2,n}$ converges to zero faster than $n^{\beta }$, with $\beta$ defined as in \eqref{sec:Rev.9-1}. The gamma error $I_\zs{1,n}$ approaches $2\min (S_\zs{1},K)$ at the same rate. On the other hand, the total trading volume $J_\zs{n}$ converges 
in probability 
to the random variable
$J({S}_\zs{1},y_\zs{1},\varrho)$, defined by 
\begin{equation}\label{sec:Mar.5}
J(x,y,\varrho)=x\int_\zs{0}^{+\infty}
\,
\lambda^{-1/2}\wt{\varphi}(\lambda,x)
\,\E\,
\left|
{\sigma(y)}{\varrho}^{-1} {Z}
+q(\lambda,x)
\right| 
\,
\d\lambda
\,,
\end{equation}
where ${Z}\sim \cN(0,1)$ independent of $S_1$ and $y_\zs{1}$. Our goal is to study the convergence of the normalized replication error corrected by these explicit limiting values, by
applying the theory of limit theorems for martingales \cite{Hall}. To do so, we search for the martingale part in the approximation of the above terms by developing a special
discretization procedure in Section \ref{Proofs}.

\begin{theorem}\label{Th.sec:Mar.1}
Suppose that condition $(\C_1)$ holds and $\varrho>0$ is a fixed positive constant. Then, 
$$
n^{\beta }(V^{n}_\zs{1}-h(S_\zs{1})-\min(S_\zs{1},K) 
+\kappa_\zs{*}J({S}_\zs{1}, y_\zs{1},\varrho))
$$
weakly converges to a centered mixed Gaussian variable as $n\to\infty$.
\end{theorem}

\begin{remark}\label{Re.sec:Mar.0}
This theorem is a generalization that includes an improved convergence rate of the results 
in \cite{Kab-Saf1,Per}, where the uniform revision is taken and the volatility is assumed to be a
constant.
\end{remark}
\begin{remark}\label{Re.sec:Mar.0-1} Note that $h(x)+\min(x,K)= x$, where $h(x)=(x-K)^{+}$ is the payoff of a classical European call option. Then, from Theorem \ref{Th.sec:Mar.1}, the wealth process $V^{n}_\zs{1}$ approaches $S_1-\kappa_\zs{*}J({S}_\zs{1},y_\zs{1},\varrho)$ as $n\to\infty$. This can be explained by the fact that the option is now sold at a higher price because $
C( 0,S_\zs{0},\hat{\sigma} ) \to S_\zs{0}$ as $\hat{\sigma}\to\infty.
$
In other words, Leland's strategy now converges to the well-known buy-and-hold strategy \cite{Kar}: buy a stock share at time $t=0$ for price $S_\zs{0}$ and keep
it until expiry.
\end{remark}
%\subsection{Improvement of rate of convergence}
We now present a method for improving the rate of convergence in Theorem \ref{Th.sec:Mar.1}. To this end, by letting $\varrho \rightarrow \infty $, we
observe that
\begin{equation} \label{sec:Mar.7}
\lim_\zs{\varrho\to\infty }J(x,y,\varrho)
=x\int_\zs{0}^{+\infty }\lambda
^{-1/2}\wt{\varphi}(\lambda,x)\vert q(\lambda,x)\vert\d \lambda\,:= J^{\ast }(x),
\end{equation}
which is independent of $y$. This suggests that the rate of convergence in Theorem \ref{Th.sec:Mar.1} can be improved if $\varrho$ is taken as a function of $n$. Our next result is established under the following condition on $\varrho$.

\noindent$(\C_\zs{2})$
{\em The parameter $\varrho =\varrho (n)$ is a function of $n$ such that
$$
\lim_\zs{n\to\infty}\,\varrho (n)=\infty
\quad\mbox{and}\quad
\lim_\zs{n\to\infty}
{\varrho}\,{n^{-\frac{\mu}{2( \mu +2) }}}=0
\,.
$$
}
\begin{theorem}\label{Th.sec:Mar.2}
Under conditions $(\C_\zs{1})-(\C_\zs{2})$, 
$$
\theta_\zs{n}(V_\zs{1}^{n}
-
h(S_\zs{1})
-\min (S_\zs{1},K)
+\kappa_\zs{*}\,
J^{*}(S_\zs{1})), \quad \mbox{with}\quad \theta_\zs{n}=n^{\beta }\varrho ^{2\beta },
$$
weakly 
converges to a centered mixed Gaussian variable as $n\to\infty$.
\end{theorem}

\begin{remark}\label{Re.sec:Mar.1}
The asymptotic distributions in Theorems \ref{Th.sec:Mar.1} and \ref{Th.sec:Mar.2} are explicitly determined in the proofs in Section \ref{Proofs}.
%Furthermore, these results still hold if
%%\begin{equation}\label{sec.Mar.3}
 %$\wh{\sigma}_\zs{t}^{2}=\sigma^2(y_\zs{t})+\varrho \sigma(y_\zs{t})\sqrt{n f'(t)}$
%%\end{equation}
%and the limit of transaction costs is now given by
%\begin{equation}\label{sec:Mar.5-0}
%J^{\prime}(x,\varrho)=x\int_\zs{0}^{+\infty}
%\,
%\lambda^{-1/2}\wt{\varphi}(\lambda,x)
%\,\E\,
%\left|
%{Z}{\varrho}^{-1} 
%+q(\lambda,x)
%\right| 
%\,
%\d\lambda
%.
%\end{equation}
%However, such a use for enlarged volatility is far away from practical significance as discussed in Subsection \ref{adjform}.
\end{remark}

\subsection{Asymptotic result for L\'epinette's strategy}
Let us study the replication error of L\'epinette's strategy $\ov{\gamma}_\zs{t}^{n}$, as defined in \eqref{sec:Rev.9-2}. As before, 
% \begin{equation*}\label{sec:Lepi.1}
% \ov{\gamma}_\zs{t}^{n}=\sum_\zs{i=1}^{n}
% \left(\wh{C}_\zs{x}(t_\zs{i-1},S_\zs{t_\zs{i-1}}) -
% \int_\zs{0}^{t_\zs{i-1}}\wh{C}_\zs{xt}(u,S_\zs{u})\d u
% \right) 
% \Chi_\zs{(t_\zs{i-1},t_\zs{i}]}(t)\,.
% \end{equation*}
the replicating portfolio is  
$
\ov{V}^{n}_\zs{1}=\ov{V}^{n}_\zs{0}+\int_\zs{0}^1\ov{\gamma}^n_\zs{t} \d S_\zs{t}-\kappa_\zs{*}\ov{J}_\zs{n},
$
where 
\begin{equation}\label{sec:Lepi.2}
\ov{J}_\zs{n}=\sum^{n}_\zs{i=1}\,S_\zs{t_\zs{i}}\vert \ov{\gamma}_\zs{t_\zs{i}}^{n}-\ov{\gamma}
_\zs{t_\zs{i-1}}^{n}\vert\,.
\end{equation}
Now, by It\^{o}'s formula, the tracking error is
\begin{equation}\label{sec:Mar.4-1}
\ov{V}^{n}_\zs{1}-h(S_\zs{1})=
\frac{1}{2}I_\zs{1,n}+\ov{I}_\zs{2,n} -\kappa_\zs{*}\ov{J}_\zs{n}\,,
\end{equation}
where
$\ov{I}_\zs{2,n}={I}_\zs{2,n}+\sum_{i\ge 1}(S_\zs{t_i}-S_\zs{t_{i-1}})
\int_\zs{0}^{t_{i-1}}\wh{C}_\zs{xt}(u,S_\zs{u})\d u$.
Then, we have the following strengthening of Theorem \ref{Th.sec:Rev.6}.
\begin{theorem}\label{Th.sec:Lepi.1}
Suppose that $(\C_1)$ is fulfilled. Then, for any $\varrho>0$, the sequence
$$
n^{\beta }(\ov{V}^{n}_\zs{1}-h(S_\zs{1})-\eta\min(S_\zs{1},K)),
\quad \mbox{with}\quad 
\eta=1-\kappa_\zs{*}{\sigma(y_1)}{\varrho}^{-1}\sqrt{8/\pi},
$$
weakly converges to a centered mixed Gaussian variable as $n\to\infty$.

\end{theorem}

\begin{remark}\label{Re.sec:Lepi.1}
Theorem \ref{Th.sec:Rev.6} can be established from Theorem \ref{Th.sec:Lepi.1} with $\varrho=\kappa_\zs{*}{\sigma}\sqrt{8/\pi}$ when the volatility is a constant. In addition, in our model, the parameter $\mu$ 
takes its values in the interval $[1\,,2)$, 
which is slightly more general than the condition imposed 
in Theorem \ref{Th.sec:Rev.6}. Moreover, if the classical form of adjusted volatility is applied for L\'epinette's strategy, then complete replication can be reached by taking $\varrho=\kappa_\zs{*}\sqrt{8/\pi}$, and we again have the result established in \cite{Darses}.
\end{remark}

%In the context of condition $(\C_\zs{2})$, the cumulated cost $\kappa_\zs{*}\ov{J}_\zs{n}$ converges to $0$ while the hedging error approaches to the terminal value $S_\zs{1}$ of the buy-and-hold strategy. Hence, the option is over replicated in this case, see Remark \ref{Re.sec:Mar.0-1}.

\begin{corollary}\label{Cor.sec:Lepi.1}
Under conditions  $(\C_1)-(\C_\zs{2})$, the wealth sequence $\ov{V}^{n}_\zs{1}$ converges in probability to $h(S_\zs{1})+\min(S_\zs{1},K)=S_1$.
\end{corollary}
\noindent Note that we do not obtain an improved convergence version of Theorem \ref{Th.sec:Lepi.1} because $\kappa_\zs{*}\ov{J}_\zs{n}$ converges to zero at the order of $\varrho$.

\section{Application to the pricing problem}\label{sec:Ap}
In this section, we present an application to the problem of option 
pricing with transaction costs. We first emphasize that it is
impossible to obtain a \emph{non-trivial} perfect hedge in the presence of
transaction costs, even in constant volatility models. In fact, the seller can take the 
\emph{buy-and-hold} strategy, but this leads to a high option price.
We show below that the price can be reduced in certain ways so that the payoff is
covered with a given probability.

\subsection{Super-hedging with transaction costs}
To be on the safe side, the investor searches for
strategies with terminal values greater than the payoff. Such
strategies are solutions to dynamic optimization problems. More precisely, 
let $H$ be a general contingent claim and let $\mathcal{A(}x)$ and $V_\zs{T}^{\pi,x}$ be the set of all 
\emph{admissible strategies} $\pi $ with initial capital $x$ and the
terminal value of strategy $\pi $, respectively. Then, the
super-replication cost for $H$ is determined as
\begin{equation}\label{sec:Ap.1}
U_\zs{0}=\inf \left\{ x\in \mathbb{R}: \exists \pi \in A(x),\, V_\zs{T}^{\pi ,x}\geq H
\quad\mbox{a.s.}\right\},
\end{equation}
(see \cite{Kar} and the
references therein for more details). In the presence of transaction costs, Cvitani\'{c} and Karatzas \cite{Cvitanic} show
that the \emph{buy-and-hold strategy} is the unique choice for super-replication, and then $S_\zs{0}$ is
the super-replication price. In this section, we show that this property still holds for approximate super-hedging.
The following observation is a direct consequence of Theorem \ref{Th.sec:Mar.2} when $\varrho$ is a function of $n$.
\begin{proposition}\label{Pr.Apl1}
Under conditions $(\C_1)-(\C_\zs{2})$, $\P-\lim_\zs{n\to\infty}V^n_1\ge h(S_1)$. The same property holds for L\'epinette's strategy.
\end{proposition}
\proof
Note first that $J^*(x)\le \min(x,K)$, for all $x>0$. Hence, by Theorem \ref{Th.sec:Mar.2},
\begin{equation}\label{sec:Ap.05}
\P-\lim_\zs{n\to\infty}(V^n_1-h(S_1))\ge (1-\kappa_\zs{*})\min(S_1,K).
\end{equation}
The left-hand side is obviously non-negative as $\kappa_\zs{*}<1$. The conclusion follows from Theorem \ref{Th.sec:Lepi.1}.\endproof

\subsection{Asymptotic quantile pricing}\label{Redu}

As seen ealier, super-hedging in the presence of transaction costs leads to a high option price. Practically, one can ask by how much the initial capital can be reduced in exchange for a shortfall probability at the terminal moment. More precisely, for a given significance level $0\leq \varepsilon \leq 1$, the seller may look for hedges with a minimal initial cost %
$$
\inf \left\{ x\in \mathbb{R}, \exists \pi \in A(x):\P\left( V_\zs{T}^{\pi ,x}\geq
H\right) \geq 1-\varepsilon \right\}.
$$%
This construction is motivated by quantile hedging theory, which goes back to \cite{Follmer,Novikov}. For related discussions, we refer to \cite{Follmer,Novikov,Per,Baran,Bratyk,Barski}.
Here, we adapt this idea to the hedging problem. Recall that the super-hedging price of Leland's algorithm is $S_0$. On the seller's side, we propose a price $\delta S_\zs{0}<S_0$ for the option, for a properly chosen $0<\delta <1$. We then follow Leland's strategy for replication. To be safe at the terminal moment, we need to choose $\varrho$ such that the probability of the terminal portfolio exceeding the sum of the real objective (i.e., the payoff) and the additional amount $(1-\delta )S_\zs{0}$ is greater than $1-\varepsilon$. Here, $\varepsilon$ is a significance level predetermined by the seller. By Proposition \ref{Pr.Apl1}, this goal can be achieved for sufficiently large $\varrho$. To determine the option price, it now remains to choose the smallest value of $\delta $. Motivated by \eqref{sec:Ap.05}, we define
this by
\begin{equation}\label{sec:Ap.5}
\delta_\zs{\varepsilon }=\inf \left\{ a>0:\Upsilon( a) \geq 1-\varepsilon
\right\}, \quad \Upsilon(a) =\P\left((1- 
\kappa_\zs{*}) \min (S_\zs{1},K)
>( 1-a) S_\zs{0}\right).
\end{equation}
Thus, the reduction in the option price is given by $(1-\delta_\zs{\varepsilon })S_0$. Clearly, smaller values of $\delta_\zs{\varepsilon }$  yield cheaper options.

Next, we show that the option price is significantly reduced, compared with powers of the parameter $\varepsilon $.
\begin{proposition}\label{Pr.sec:Ap.1}
Assume that $
\sigma_\zs{\max}=
\sup_\zs{y\in\bbr}\sigma (y)<\infty\,.
$
Then, for any 
$r>0$ and $\delta_\zs{\varepsilon }$ defined by 
\eqref{sec:Ap.5},

\begin{equation}\label{sec:Ap.5-0}
\lim_\zs{\varepsilon \rightarrow 0}
({1-\delta_\zs{\varepsilon }}){\varepsilon^{-r}}=+\infty\,.
\end{equation}
\end{proposition}
\proof
We first observe that $0<\delta_\zs{\varepsilon }\leq 1$ and $\delta_\zs{\varepsilon }$ 
tends to $1$ as $\varepsilon \rightarrow 0$. Set
$b=1-\kappa_\zs{*}$. Then, for sufficiently small $\varepsilon$ such that 
$\delta_\varepsilon >a> 1-bK/S_\zs{0}$, one has
\begin{align*}
1-\varepsilon 
&> \P(b \min(S_\zs{1},K)>(1-a) S_\zs{0}) =1-\P(S_\zs{1}/S_\zs{0}\leq ( 1-a) /b) .
\end{align*}
Therefore, 
\begin{equation}\label{sec:Ap.5-1}
  \varepsilon 
< \P\left(S_\zs{1}/S_\zs{0}\le (1- a)/b\right)
\leq \P\left( 
X_\zs{1}
\le -z_\zs{a}\right),
 \end{equation}
where $X_\zs{t}=\int_\zs{0}^{t}\sigma (
y_\zs{t}) \d W_\zs{t}^{(1)}$ and 
$z_\zs{a}=\ln( b/(1-{a}))-\sigma_\zs{\max}^{2}/2$.
To estimate this probability, we note  that 
for any integer $m\ge 1$,
$
\E\left(X_\zs{1}\right)^{2m}
\leq \sigma_\zs{\max}^{2m}(2m-1)!!
$
(see \cite[Lemma 4.11, p.130]{Lip-Shir}).
Set $R( \upsilon )=2 \upsilon  \sigma_\zs{\max}^{2}$. For any $0<\upsilon <1/2\sigma_\zs{\max}^{2}$,
\begin{align*}
\E
\,e^{ \upsilon  X_\zs{1}^{2}}=\sum_\zs{m=0}^{\infty }\frac{ \upsilon  ^{m}}{m!
}\E\left( X_\zs{1}\right)^{2m} \leq\sum_\zs{m=0}^{\infty }\frac{ \upsilon  ^{m}}{m!}\sigma_\zs{\max}^{2m}( 2m-1)
!!
\leq \,\frac{1}{1-R( \upsilon )}\,.
\end{align*}
Therefore, for sufficiently small $\varepsilon>0$, we have
$$
\varepsilon\leq\P( X_\zs{1}\leq -z_\zs{a}) =\P(
-X_\zs{1}\geq z_\zs{a}) \leq e^{- \upsilon  z^{2}_\zs{a}}\,\E
\,e^{ \upsilon  X_\zs{1}^{2}}\leq 
\frac{e^{- \upsilon  z^{2}_\zs{a}}}{1-R( \upsilon )}.
$$
Then, $1-a\geq b\,e^{-\iota_\zs{\varepsilon}(\upsilon)}$, where
$\iota_\zs{\varepsilon}(\upsilon)=\sqrt{
\left\vert \ln \varepsilon ( 1-R( \upsilon ) )
\right\vert/ \upsilon}+\sigma_\zs{\max}^{2}/2$.
Letting $a\to\delta_\zs{\varepsilon }$, we get
$1-\delta_\zs{\varepsilon }\geq b e^{-\iota_\zs{\varepsilon}(\upsilon)}$,
which implies
 \eqref{sec:Ap.5-0}.\endproof

\vspace{1mm}
The boundedness of the volatility function is essential for the above comparison result. If we wish to relax this assumption, the price reduction will be smaller than that in Proposition \ref{Pr.sec:Ap.1}. 

\begin{proposition}\label{Pr.sec:Ap.2}
 Suppose that $\E \exp\{\alpha\int_\zs{0}^1\sigma^2(y_s)ds\}<\infty$, for some constant $\alpha>1/2$. Then, for
$r_{\alpha}=(2\sqrt{2\alpha}+1)/2\alpha$,
\begin{equation}\label{sec:Ap.6}
\liminf_\zs{\varepsilon \rightarrow 0}\,\varepsilon^{-r_{\alpha}}
\left(1-\delta_\zs{\varepsilon}\right)>0\,.
\end{equation}
\end{proposition}

\proof 
For any positive constant $L$ we set
\begin{equation}
\tau=
 \tau_\zs{L}=\inf\left\{t>0:\int_\zs{0}^t\sigma^2(y_s)\d s\ge L\right\}\wedge 1,
\end{equation}
which is understood to be the first time that the log-price's variance passes level $L$. Then, from \eqref{sec:Ap.5-1},
\begin{align}\label{sec:Ap.7}
\varepsilon\le 
\P\left(
\cE^{-1}_\zs{1}(\sigma)
\ge u_\zs{a}, \int_\zs{0}^{1}\sigma^2(y_s)\d s\le L \right)
+\P\left(\int_\zs{0}^{1}\sigma^2(y_s)\d s\ge L\right),
\end{align}
where 
$\cE_\zs{t}(\sigma)=e^{\int_\zs{0}^{t}\,\sigma(y_s)\d W^{(1)}_\zs{s}
-\frac{1}{2}\int_\zs{0}^{t}\sigma^2(y_s)\d s}$,
$u_\zs{a}=(1-\kappa_\zs{*})/(1-a)$,
and
$\delta_\varepsilon >a> 1-bK/S_\zs{0}$. 
Note that for any $p>0$,  the stopped process 
$\chi_\zs{t}=\cE_\zs{\tau\wedge t}(-p\sigma)$
is a martingale and $\E \chi_\zs{t}=1$.
Therefore, the first probability on the right side of \eqref{sec:Ap.7} 
can be estimated as
$$
(u_\zs{a})^{-p}\,
\E\,\cE^{-p}_\zs{\tau}(\sigma)
=(u_\zs{a})^{-p}\,\E\,\chi_\zs{1}\,
e^{\check{p}\int_\zs{0}^{\tau}\,\sigma^2(y_s)\d s}
\le (u_\zs{a})^{-p}\,e^{\check{p}L}\,,
$$
where $\check{p}=(p^2+p)/2$. 
%For $p^2\le 2\alpha$, 
By the hypothesis and Chebysev's inequality, we have 
$$
\P\left(\int_\zs{0}^{1}\sigma^2(y_s)\d s\ge L\right)\le C_\zs{\alpha} e^{-\alpha L},
\quad\mbox{with}\quad  C_\zs{\alpha}=\E \exp\left\{\alpha\int_\zs{0}^1\sigma^2(y_s)\d s\right\}.
$$ 
Hence,
$
\varepsilon\le(u_\zs{a})^{-p}\,e^{\check{p}L}+C_\zs{\alpha} e^{-\alpha L}.
$
By choosing $L=\alpha^{-1}\ln(2C/\varepsilon)$
and letting $a\to \delta_{\varepsilon}$, one deduces 
that for any $p>0$ and for some positive constant $\tilde{C}_\zs{\alpha}$,
$$
1-\delta_{\varepsilon}\ge \tilde{C}_\zs{\alpha}\,{\varepsilon}^{\gamma^{*}(p)},
\quad\mbox{where}\quad
\gamma^{*}(p)=(p+1)/(2\alpha)+p^{-1}\,.
$$
Note that 
$r_{\alpha}=\min_\zs{p>0} \gamma^{*}(p)=  \gamma^{*}(\sqrt{2\alpha})$.
Then, including in the last inequality $p=\sqrt{2\alpha}$
we obtain property \eqref{sec:Ap.6}.\endproof
\begin{remark}\label{Re.sec:App.2}
It is clear that $r_\zs{\alpha}<1$ for $\alpha>3/2+\sqrt{2}$.
 The condition used in  Proposition~\ref{Pr.sec:Ap.2} holds
%$\E \exp\{\alpha\int_\zs{0}^1\sigma^2(y_s)ds\}<\infty$ 
for such $\alpha$,  when $\sigma$ is linearly bounded and $y_\zs{t}$ follows an Orstein-Uhlenbeck process (see the Appendix \ref{OU}).
The same quantile pricing result can be established for the L\'epinette strategy.
\end{remark}

\section{Examples and numerical results}\label{Exa}
In this section, we list some well-known SV models for which condition $(\C_1)$ is fulfilled. To this end, we need some moment estimates for solutions to general SDEs, 
\begin{equation}\label{sec:Exa.1-0}
 \d y_\zs{t}=F_1(t,y_\zs{t})\d t+F_2(t,y_\zs{t})\d Z_\zs{t},\quad y(0)=y_0,
\end{equation}
where $Z$ is a standard Wiener process and $F_1, F_2$ are two smooth functions. We first recall the well-known result in SDE theory (see for example \cite[Theorem 2.3, p.107]{Fried}).

\begin{theorem}\label{Th.sec:Exa.1}
Suppose that $F_1(t,y)$ and $\,F_2(t,y)$ are measurable in $(t,y)\in [0,T]\times \bbr$, linearly bounded and locally Lipschitz. If $\E\,\vert y_\zs{0}\vert^{2m}<\infty$ for some integer $m\ge1$, then there exists a unique solution $(y_\zs{t})$ to \eqref{sec:Exa.1-0} and
$$
\E\, \vert y_\zs{t}\vert ^{2m}<(1+\E\,\vert y_\zs{0}\vert^{2m})e^{\alpha t},\,\quad \E \sup_\zs{0\le s\le t}\vert y_s\vert ^{2m}<M(1+\E\,\vert y_\zs{0}\vert^{2m}),
$$
where $\alpha,M$ are positive constants dependent on $t,m$.
\end{theorem}
\noindent In the context of Theorem \ref{Th.sec:Exa.1}, condition $(\C_1)$ holds if the volatility function and its derivative have polynomial growth $\vert\sigma(y)\vert \le C(1+\vert y\vert^m)$, for some positive constant $C$ and $m\ge1$.

\vspace{1mm}
\noindent{\bf Hull-White models:}
Assume that $y_\zs{t}$ follows a geometric Brownian motion
\begin{equation}\label{sec:Exa.0}
\d S_\zs{t}=(y_\zs{t}+\sigma_\zs{\min})S_\zs{t}\d W_\zs{t} \quad \mbox{and}\quad
\d y_\zs{t}=y_\zs{t}(a\d t+b\d Z_\zs{t}),
\end{equation}
where $\sigma_\zs{\min}>0$, $a$ and $b$ are some constants, and $Z$ is a standard Brownian motion correlated with $W_\zs{t}$.
Put $y^*= \sup_\zs{0\le t\le 1} \vert y_\zs{t}\vert $. Then, by Theorem \ref{Th.sec:Exa.1}, we have
$$
\E \,(y^*)^{2m} \le C(1+\E \vert y_\zs{0}\vert^{2m})<\infty, \quad m\ge 1,
$$
as long as $\E \vert y_\zs{0}\vert^{2m}<\infty$.
Therefore, condition $(\C_1)$ is fulfilled in \eqref{sec:Exa.0}.

\vspace{1mm}
\noindent{\bf Uniform elliptic volatility models:} Suppose that volatility is driven by a mean-reverting Orstein-Uhlenbeck process
\begin{equation}\label{sec:Exa.2}
\d S_\zs{t}=(y_\zs{t}^2+\sigma_\zs{\min})S_\zs{t}\d W_\zs{t}\quad \mbox{and}\quad 
\d y_\zs{t}=(a-by_\zs{t})\d t+ \d Z.
\end{equation}
In this case,  $\sigma(y)=y^2 +\sigma_\zs{\min}$. Thus, condition $(\C_1)$ is verified throughout Theorem \ref{Th.sec:Exa.1}.

\vspace{1mm}
\noindent{\bf Stein-Stein models:} Assume that
\begin{equation}\label{sec:Exa.3}
\d S_\zs{t}=\sqrt{y^2_\zs{t}+\sigma_\zs{\min}}\,S_\zs{t}\d W_\zs{t}\quad \mbox{and}\quad
\d y_\zs{t}=(a-by_\zs{t})\d t+\d Z_\zs{t}.
\end{equation}
We have $\sigma(y)=\sqrt{y^2+\sigma_\zs{\min}}$ and condition $(\C_1)$ is also verified by Theorem \ref{Th.sec:Exa.1}.

\vspace{1mm}
\noindent{\bf Heston models:} Heston \cite{Heston} proposes a SV model where volatility is driven by a CIR process, which also known as a square root process. This model can be used in our context. 
Indeed, assume now that the price dynamics are given by the following:
\begin{equation}\label{sec:Exa.1.1}
\d S_\zs{t}=\sqrt{ y_\zs{t}+\sigma_\zs{\min}}\,S_\zs{t}\d W_\zs{t}\quad \mbox{and}\quad
\d y_\zs{t}=(a-b y_\zs{t})\d t+\sqrt{y_\zs{t}}\,\d Z_\zs{t},\quad y_\zs{0}\ge 0.
\end{equation}
For any $a$ and $b>0$, the last equation admits a unique strong solution $y_\zs{t}>0$. Note that the Lipschitz condition in Theorem \ref{Th.sec:Exa.1} is violated, but by using the stopping times technique, we can directly show that $\E\,y^*<\infty$. %(see the Appendix \ref{Ap:Heston}) 
Hence, this implies that condition $(\C_1)$ is satisfied for model \eqref{sec:Exa.1.1}.

Similarly, we can check that $(\C_1)$ also holds for Ball-Roma models \cite{Bal-Rom} or, more generally, for a class of processes with bounded diffusion satisfying the following condition.  

\vspace{1mm}
{\em
\noindent $(\A)$ There exist positive constants $a,b,$ and $M$ such that
$$yF_1(t,y)\le a-b y^2\quad\mbox{and}\quad \vert F_2(t,y)\vert\le M,\quad \mbox{for all}\quad t>0, y\in\bbr.$$
}
\begin{proposition}\label{Pr.sec:Exa.2}
 Under condition $(\A)$, there exists $\alpha>0$ such that $\E e^{\alpha y^{*2}}<\infty$.
\end{proposition}
\proof The proof uses the same method as in Proposition 1.1.2 in \cite{Kab-Ser}. \endproof

\vspace{1mm}
\noindent{\bf Scott models}: Suppose that volatility follows an Orstein-Uhlenbeck, as in Stein-Stein models, and the function $\sigma$ takes the exponential form
\begin{equation}\label{sec:Exa.1}
\d S_\zs{t}=(e^{\delta y_\zs{t}}+\sigma_\zs{\min})S_\zs{t}\d W_\zs{t}^{(1)}\quad
\mbox{and}\quad
\d y_\zs{t}=(a-by_\zs{t})\d t+\d Z_\zs{t},
\end{equation}
where $a,b$ and $\sigma_\zs{\min}>0$ are constants. Here, $\delta>0$ is chosen such that $2\delta\le\alpha$, defined as in Proposition \ref{Pr.sec:Exa.2}. Clearly, $\sigma(y)=e^{\delta y}+\sigma_\zs{\min}$ and condition $(\C_1)$ is fulfilled because 
$$
\E\sup_\zs{0\le t\le 1}\vert \sigma(y)\vert^2 \le 2\sigma_\zs{\min}^2+2\E\, (e^{2\delta}{\bf 1}_\zs{\left\{\vert y_\zs{t}\vert \le 1\right\}}+e^{2\delta\vert y\vert_1^2}{\bf 1}_\zs{\left\{\vert y_\zs{t}\vert > 1\right\}})<\infty.
$$

%\noindent{\large \bf Appendix: Proof  of Main Results}
\noindent{\bf Numerical result for the Hull-White model}: We provide a numerical example for the approximation result of L\'epinette's strategy in the Hull-White model \eqref{sec:Exa.0}. By Theorem \ref{Th.sec:Lepi.1}, the  corrected replication error is given by
$V_1^n-\max(S_1-K,0)-\eta\min(S_1,K)$, where $\eta=1-\kappa_\zs{*}{\sigma(y_1)}{\varrho}^{-1}\sqrt{8/\pi}$. 
%\begin{figure}[h]
%\centering
%\includegraphics[width=0.8\columnwidth,height=11cm]{StockVolPaths.png}%
%\caption{Paths of stock price and volatility of Hull-White model}%$\sigma_{\min}=1, a=-2, b=1$.}%
%\label{fig.Ch_2.SVStockVolPath}%
%\end{figure}
\begin{table}[h]\begin{center}{\small
\begin{tabular}{|l|l|l|l|l|l|l|l|}
\cline{1-7}
\multicolumn{1}{|c|}{$n$}&\multicolumn{1}{|c|}{gain/loss}&\multicolumn{1}{|c|}{corrected error}&\multicolumn{1}{|c|}{lower bound}&\multicolumn{1}{|c|}{upper bound}&\multicolumn{1}{|c|}{price}& \multicolumn{1}{|c|}{strategy}\\
\cline{1-7}
\hline
10 & 0.1523845 & -0.2225988 & -0.2363122 & -0.2088854 & 0.7914033 & 0.9013901  \\ 
\hline
50 & 0.2966983 & -0.0596194 & -0.0670452 & -0.0521936 & 0.9399330 & 0.9706068  \\ 
\hline
100 & 0.3086120 & -0.0288526 & -0.0350141 & -0.0226911 & 0.9746527 & 0.9875094  \\ 
\hline
500 & 0.2955755 & 0.0032387 & -0.0005821 & 0.0070594 & 0.9991733 & 0.9995891  \\ 
\hline
1000 & 0.2851002 & 0.0012409 & -0.0021596 & 0.0046415 & 0.9999300 & 0.9999652  \\ 
\hline
\end{tabular}
}
\caption{\small Convergence for L\'epinette's strategy with $\kappa_\zs{*}=0.01, \varrho=2.$ }\label{Tab.1}
\end{center}\end{table}
\begin{table}[h]\begin{center}{\small
\begin{tabular}{|l|l|l|l|l|l|l|l|}
\cline{1-7}
\multicolumn{1}{|c|}{$n$}&\multicolumn{1}{|c|}{gain/loss}&\multicolumn{1}{|c|}{corrected error}&\multicolumn{1}{|c|}{lower bound}&\multicolumn{1}{|c|}{upper bound}&\multicolumn{1}{|c|}{price}& \multicolumn{1}{|c|}{strategy}\\
\cline{1-7}
10 & 0.2859197 & -0.0744180 & -0.0813544 & -0.0674816 & 0.9246420 & 0.9659700  \\ 
\cline{1-7}
50 & 0.3172523 & -0.0069238 & -0.0115426 & -0.0023049 & 0.9921661 & 0.9962377  \\ 
\cline{1-7}
100 & 0.3033519 & 0.0007474 & -0.0030916 & 0.0045864 & 0.9984346 & 0.9992385  \\ 
\cline{1-7}
500 & 0.3618707 & 0.0001296 & -0.0024741 & 0.0027333 & 0.9999977 & 0.9999989  \\ 
\cline{1-7}
1000 & 0.3334375 & 0.0003996 & -0.0020559 & 0.0028550 & 1 & 1 \\ 
\cline{1-7}
\end{tabular}
}
\caption{\small Convergence for L\'epinette's strategy with $\kappa_\zs{*}=0.001, \varrho=4$.}\label{Tab.2}
\end{center}\end{table}
The difference $V_1^n-\max(S_1-K,0)$ can be seen as the gain/loss of strategy $\bar{\gamma}^n$. For a numerical evaluation, we simulate $N=500$ trajectories in a crude Monte-Carlo method, where the correlation coefficient of the two Brownian motions is $0.05$ and the other initial values are $S_0=K=1,\, y_\zs{0}=2$, $\sigma_{\min }=2, a=-2$ and $b=1$. For each value of $n$, we estimate the average value of the corrected error and give the corresponding $95\%$ intervals defined by lower and upper bounds. Initial numbers of shares held are given in the last column of Tables \ref{Tab.1} and \ref{Tab.2}. It turns out that strategy $\bar{\gamma}_t^n$ converges to the buy-and-hold strategy and the option prices approach the super-hedging price $S_0$. We also see that the convergence of the corrected replication error to zero is somehow slow. In fact, increasing values of $\varrho$ can provide a faster convergence, but this unexpectedly leads to super-replication more rapidly. 

\begin{center}  
\begin{figure}[h]
\begin{subfigure}{0.52\textwidth}
\includegraphics[width=\columnwidth,height=6cm]{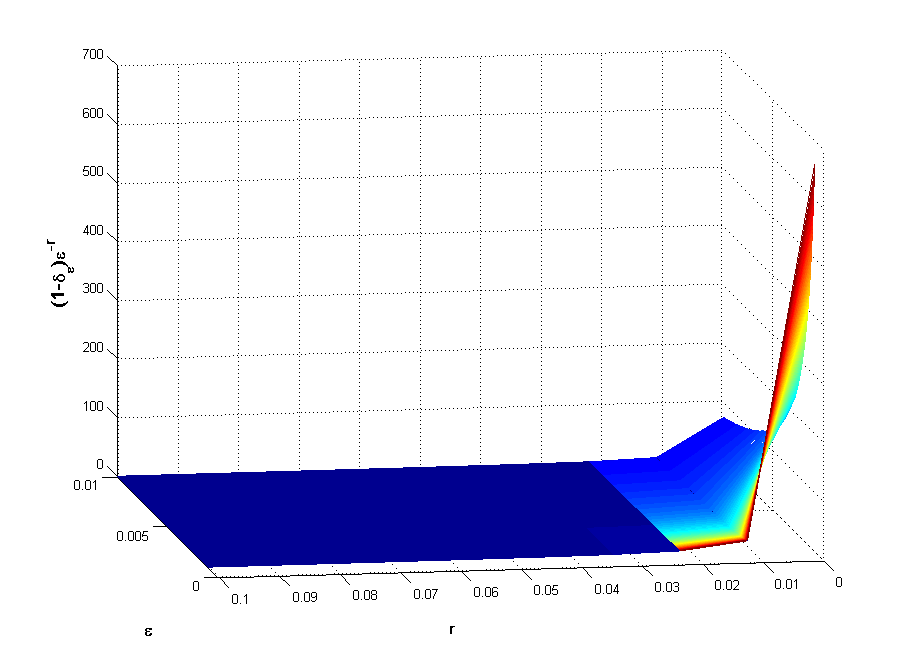}
\caption{\small Reduction factor $1-\delta_\varepsilon$ and powers of $\varepsilon$}%
\label{Ch_7.SH.surfdelta}
\end{subfigure}
\begin{subfigure}{0.52\textwidth}
\includegraphics[width=\columnwidth,height=6cm]{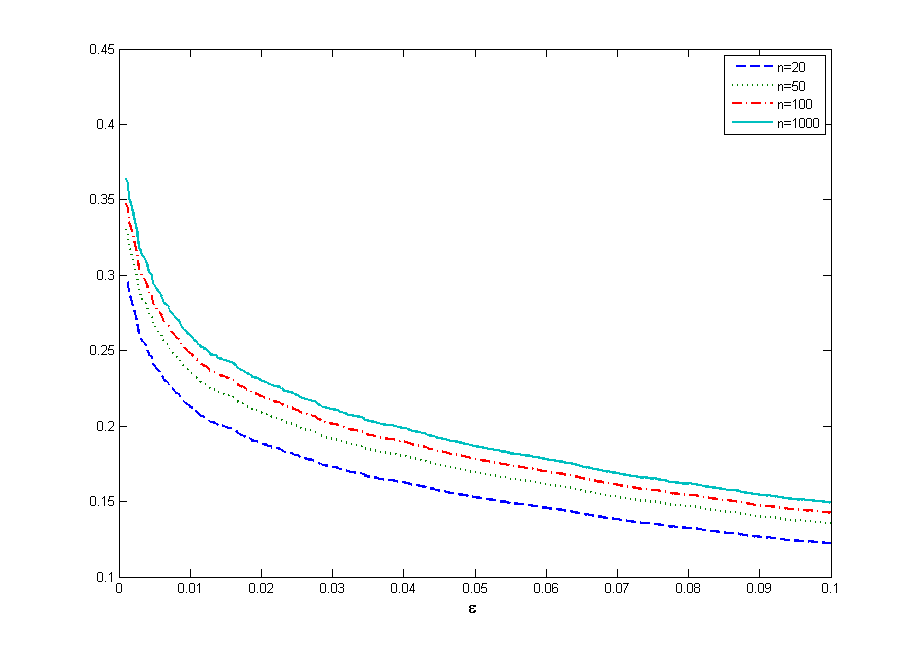}
\caption{\small Reduction amount and $\varepsilon$}%%
\label{Ch_7.SH.red2}
\end{subfigure}
\end{figure}\end{center}

We now provide a numerical illustration for the quantile hedging result of Proposition \ref{Pr.sec:Ap.1}. For simplicity, suppose that $\sigma(y)=\sin^2 (y)+0.1$ and that $y$ follows a geometric Brownian motion as above. To compare the reduction factor $1-\delta_\varepsilon$ with powers of significance level $\varepsilon$, we compute $(1-\delta_\varepsilon)\varepsilon^{-r}$ for $0.001\le \varepsilon \le 0.1$ and $0\le r \le 0.1$, with $\kappa_*=0.001$. Then, \eqref{sec:Ap.5-0} is confirmed by the simulation result (see Figure \ref{Ch_7.SH.surfdelta}). The simulation also shows that the option price inclusive of transaction costs is $1-0.385=0.615$, which is cheaper than the super-hedging price $S_0=1$, for a shortfall probability less than $0,1\%$.
Of course, it is reasonable to replace $S_0$ by the option price inclusive of transaction costs $\wh{C}(0,S_0)$. The simulated reduction in the option price $(1-\delta_\varepsilon)\wh{C}(0,S_0)$ is then given in Figure \ref{Ch_7.SH.red2}.
%\begin{figure}%
%\includegraphics[width=\columnwidth, height=7cm]{QuantilePrice.png}%
%\caption{Reduced price and reduction amount}%
%\label{Ch_7.SH.red1}%
%\end{figure}

\section{High-frequency markets}\label{hfm}
We now assume that purchases of the risky asset are carried out at a higher ask price $S_\zs{t}+\varepsilon_\zs{t}$, whereas sales earn a lower bid price $S_\zs{t}-\varepsilon_\zs{t}$. Here the mid-price $S_\zs{t}$ is given as in model \eqref{sec:Mar.1} and $\varepsilon_\zs{t}$ is the halfwidth of the bid-ask spread. Then, for any trading strategy of finite variation $\psi_\zs{t}$, the wealth process can be determined by
\begin{equation}\label{hfc0}
V_\zs{t}=V_0+\int_\zs{0}^t \psi_s\d S_s - \int_\zs{0}^t \varepsilon_s \d \vert \psi\vert_s,
\end{equation}
where $\vert \psi\vert$ is the total variation of $\psi_\zs{t}$. Observe that the first two terms are the classic components in frictionless frameworks, and respectively describe the initial capital and gains from trading. The last integral in \eqref{hfc0} accounts for transaction costs incurred from the trading activities by weighting the total variation\footnote{It is important to know that the classical Black-Scholes strategy is not finite variation.} of the strategy with the halfwidth of the spread. 

For optimal investment and consumption with small transaction costs \cite{Kallsen}, the additional terms should be added in the formulation of $V_\zs{t}$. In such cases, approximate solutions are usually determined through an asymptotic expansion around zero of the halfwidth spread $\varepsilon$, where the leading corrections are obtained by collecting the inputs from the frictionless problem. 

In this section, we are only interested in replication using discrete strategies in Leland's spirit. Assume that for replication, the option seller applies a discrete hedging strategy $\psi_\zs{t}^{n,\varepsilon}$, revised at $n$ dates defined by $t_i=g(i/n)$ as in Section \ref{Mar}. The corresponding wealth process is now given by
\begin{equation}\label{hfc1}
V_\zs{t}^{n,\varepsilon}=V_0^{n,\varepsilon}+\int_\zs{0}^t \psi_s^{n,\varepsilon}\d S_s - \sum_{i=1}^n\varepsilon_{t_i}\vert \psi^{n,\varepsilon}_{t_i}-\psi^{n,\varepsilon}_{t_{i-1}}\vert.
\end{equation}
To treat the risk of transaction costs, we again apply the increasing volatility principle. Note that in high frequency markets, the bid-ask spread is, in general, of the same order of magnitude as price jumps\footnote{We thank an anonymous referee for pointing out the correspondence of the case $\alpha=1/2$ to this setting.} . Hence, $\varepsilon_\zs{t}$ should be of the form $\kappa_\zs{*} n^{-1/2}S_\zs{t},$ for some positive constant $\kappa_\zs{*}$. Then, this case corresponds to the Leland-Lott framework with $\alpha=1/2$.

In our context, the method in Section \ref{Mar} is still helpful when $\psi_\zs{t}^{n,\varepsilon}$ is replaced by Leland's or L\'epinette's strategy. 

% \begin{remark}
% Recall that when $\alpha=1/2$, $\wh{\sigma}$ is independent of $n$. For this case, it would require some modification in our approximation procedure where the property $\wh{\sigma}\to\infty$ plays the central role. However, in stochastic volatility models, it is not clear that the Kabanov-L\'epinette's theorem still holds and the dependence on the true volatility of the classical form may cause problems in a practical point of view.
% \end{remark}

% \vspace{2mm}
% Interestingly, the choice \footnote{We take the uniform revision cas for simplicity.} $\wh{\sigma}^2=\varrho(n)\to\infty$  can be made to get an approximately complete replication as in Corrollary \ref{Cor.sec:Lepi.1}. Here the rate of convergence for this case is order of $\varrho$.
\begin{proposition}
Let $\varepsilon_\zs{t}=\kappa_\zs{*} n^{-1/2}S_\zs{t}$, and assume that the adjusted volatility is of the form $\wh{\sigma}^2=\varrho\sqrt{n f'(t)}$ as in \eqref{sec:Mar.2}. For both Leland's and L\'epinette's strategies, the sequence of replicating portfolio values ${V}^{n,\varepsilon}_\zs{1}$ converges in probability to $h(S_\zs{1})+\min(S_\zs{1},K)=S_1$. In particular, $n^\beta({V}^{n,\varepsilon}_\zs{1}-S_1)$ converges to a mixed Gaussian variable as $n\to\infty$.
\end{proposition}

\proof The proof is a direct consequence of Theorem \ref{Th.sec:Mar.1}, because the total transaction cost now converges to zero. \endproof

Note that the case $\alpha=0$ studied in Section \ref{Mar} corresponds to the assumption $\varepsilon_\zs{t}=\kappa_\zs{*}S_\zs{t}$, for some constant $\kappa_*$. This specific form means that the market is more illiquid and the bid-ask spread is now proportional to the spot price in every trade. Therefore, approximate hedging results for this case are the same as those in Section \ref{Mar}.

We conclude the section by supposing that the stock spreads remain constant at all times, regardless of the current stock price. In other words, $\varepsilon_\zs{t}=\kappa_\zs{*}$ for some positive constant $\kappa_*$. Intuitively, transaction costs are now based on the volume of traded shares, instead of the traded amount of money as in the literature and Section \ref{Mar}. It is interesting to see that our methodology still works in this case. The following result is just an analog of Theorem \ref{Th.sec:Mar.1}, with a small modification to the limiting value of transaction costs, defined by
\begin{equation}\label{sec:Mar.5-2}
J_0(x,y,\varrho)=\int_\zs{0}^{+\infty}
\,
\lambda^{-1/2}\wt{\varphi}(\lambda,x)
\,\E\,
\left|
{\sigma(y)}{\varrho}^{-1} {Z}
+\frac{\ln(x/K)}{2\lambda}-\frac{1}{4}
\right| 
\,
\d\lambda
\,,
\end{equation}
where
${Z}\sim \cN(0,1)$ independent of $S_1, y_1$
\begin{proposition}\label{Th.sec:hfm.1} Suppose that $\varepsilon_\zs{t}=\kappa_*>0$ and $\wh{\sigma}^2=\varrho\sqrt{n f'(t)}$. 
For Leland's strategy under condition $(\C_1)$, the sequence
$
n^{\beta }\left(V_\zs{1}^{n,\varepsilon}
-
h(S_\zs{1})
-\min (S_\zs{1},K)
+
\kappa_*\,J_0({S}_\zs{1},y_\zs{1}\varrho)\right)
$
weakly converges to a centered mixed Gaussian variable as $n\to\infty$. Furthermore, for L\'epinette's strategy, $
n^{\beta }\left(\ov{V}^{n,\varepsilon}_\zs{1}-h(S_\zs{1})-(1-\eta_0)\min(S_\zs{1},K)\right)
$
weakly converges to a centered mixed Gaussian variable, where $\eta_0={\sigma(y_\zs{1})}\varrho^{-1} S_1^{-1}\sqrt{8/\pi}$.
%\begin{equation}\label{sec:Mar.5.1}
%\eta_0={\sigma(y_\zs{1})}\varrho^{-1} S_1^{-1}\sqrt{8/\pi}.
%\end{equation}
\end{proposition}
\proof The proof is similar to that of Theorem \ref{Th.sec:Mar.1} (see Section \ref{Proofs}).\endproof
\begin{remark} When $\varrho\to\infty$ under condition $(\C_\zs{2})$, one obtains an improved-rate version of Proposition \ref{Th.sec:hfm.1}, as in Theorem \ref{Th.sec:Mar.2}. 
\end{remark}

\section{Proofs}\label{Proofs}
Our main results are proved in the following generic procedure.

\noindent {\it Step 1}: Determine the principal term of the hedging error. In particular, we will show that the gamma term $I_{1,n}$ converges to $2\min(S_1,K)$, while the cummulative transaction cost approaches $J$ defined in \eqref{sec:Mar.5}. Both convergences are at rate $\theta_\zs{n}=n^{\beta}\varrho^{2\beta}$.

\vspace{1mm}

\noindent{\it Step 2}: Represent the residual terms as discrete martingales. To this end, stochastic integrals will be discretized by following a special procedure set up in Subsection \ref{Discr}.
\vspace{1mm}

\noindent{\it Step 3}: Determine the limit distribution of the normalized replication error by applying Theorem \ref{Th.sec:Bat.0}. This result is the key tool, but we need in fact special versions adapted to our context. These will be explicitly constructed in Subsection \ref{CTL}.

%\subsection{Basic asymptotic tools}
%\subsection{Basic approximation tools}
%\subsubsection{Motivation}\label{Moti}

\subsection{Preliminary}\label{sec:Bat}

Note first that $\wh{C}(t,x)$ and its derivatives can be represented as functions of $\lambda_\zs{t}$ and $x$, where 
\begin{equation}\label{sec:Mot.1}\lambda_\zs{t}=\lambda_\zs{0}(
1-t)^\frac{1}{4\beta}:=\lambda_\zs{0}\nu(t)\quad\mbox{and}\quad\lambda_\zs{0}=\wt{\mu}\varrho\sqrt{n}.
\end{equation} 
Moreover, the function $\wt{\varphi}(\lambda,x)$, which appears in all $k$-th ($k\ge 2)$ degree derivatives of $\wh{C}$ with respect 
to $x$ and derivatives in time via 
the relation \eqref{sec:Rev.7}, is exponentially decreasing to zero when $\lambda$ tends to zero or infinity. This property motives our analysis in terms of variable $\lambda$. In particular, let us fix two functions $l_*,\, l^*$ and let $1\le m_1<m_2\le n$ be two integers such that $l_*=\lambda_\zs{0}\nu(g(m_2/n))$ and $l^*=\lambda_\zs{0}\nu(g(m_1/n))$. Then, all terms corresponding to index $j\notin [m_1,m_2]$ can be ignored at a certain order which depends on the choice of $l_*$ and $ l^*$. 
\begin{figure}[h]\begin{center}
\includegraphics[width=0.9\columnwidth,height=6cm]{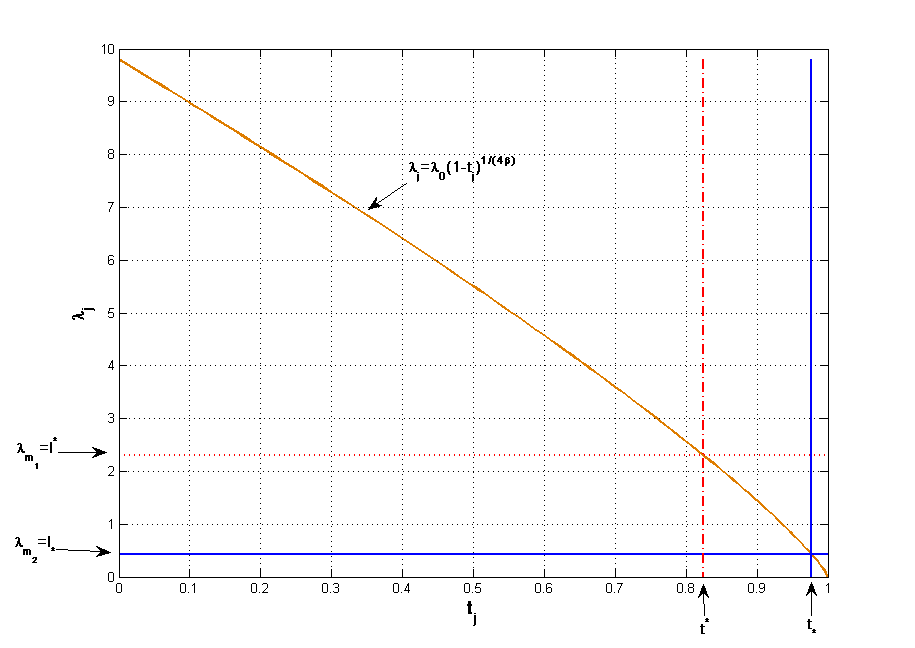}\vspace{-4mm}
\caption{The sequences $(\lambda_j)$ and $(t_j)$ defined by \eqref{sec:Bat.3}. }%
\label{Ch_2.TwoSequences}\end{center}
\end{figure}
For our purpose, the desired order is $\theta_\zs{n}\sim\lambda_\zs{0}^{2\beta}.$ Therefore, we take, for example, $
l_\zs{*}=1/\ln^{3}n,\ l^{*}=\ln^{3}n
$
and define
\begin{equation}\label{sec:Bat.1}
m_\zs{1}=n-\left[n\left( {l^{*}}/{\lambda_\zs{0}}
\right)^{2/( \mu +1) }\right]+1
\quad\mbox{and}\quad
 m_\zs{2}=n-\left[n\left({l_{*}}/{\lambda_\zs{0}}
\right)^{2/(\mu +1)}\right]\,,
\end{equation}
where the notation $[x]$ stands for the integer part of a real number $x$. Below, we focus on the subsequence $(t_j)$  of trading times 
and the corresponding sequence
$\left(\lambda_\zs{j}\right)$,
 defined as 
\begin{equation}\label{sec:Bat.3}
t_\zs{j}=1-(1-j/n) ^{\mu }
\quad\mbox{and}\quad
\lambda_\zs{j}=\lambda_\zs{0}(
1-t_\zs{j})^\frac{1}{4\beta},\quad {m_\zs{1}\le j\le m_\zs{2}}.
\end{equation}
Note that
$\left(t_\zs{j}\right)$ 
is an increasing sequence
taking values in $\left[t^{\ast},t_{\ast}\right]$, where
$t_{\ast}=1-( l_\zs{\ast}/\lambda_\zs{0}) ^{4\beta}$ and
$t^{\ast}=1-(l^{*}/\lambda_\zs{0})^{4\beta}$, whereas $\left(\lambda_\zs{j}\right)$ is decreasing in $[l_*,l^*]$.
Therefore, we use the notations $\Delta t_j= t_j-t_{j-1}$ and $\Delta \lambda_j= \lambda_{j-1}-\lambda_{j}$, for $m_1\le j\le m_2$, to avoid recopying the negative sign in discrete sums.

Below, It\^o stochastic integrals will be discretized through the following  sequences of independent normal random variables
\begin{equation}\label{sec:Tool.12-1}
{Z}_\zs{1,j}=\frac{W_\zs{t_\zs{j}}^{(1)}-W_\zs{t_\zs{j-1}}^{(1)}}
{\sqrt{t_\zs{j}-t_\zs{j-1}}}\quad \mbox{and}\quad
{Z}_\zs{2,j}=
\frac{W_\zs{t_\zs{j}}^{(2)}-W_\zs{t_\zs{j-1}}^{(2)}}{\sqrt{t_\zs{j}-t_\zs{j-1}}}.
\end{equation}
We set 
\begin{equation}\label{sec:Bat.3-1}
{p}(\lambda,x,y)
=\frac{\varrho}{\sigma(y)}\,
\left(
\frac{\ln (x/K)}{2\lambda}
-
\frac{1}{4}
\right)
\end{equation}
and write for short ${p}_{j-1}={p}(\lambda_\zs{j-1},S_\zs{t_\zs{j-1}},y_\zs{t_\zs{j-1}})$. This style of reduced notation is abusively applied for functions appearing in the approximation procedure. Define
\begin{equation}
\begin{cases}
 {Z}_\zs{3,j}=
\vert {Z}_\zs{1,j}+{p}_{j-1}\vert-
\E\,\left(\vert {Z}_\zs{1,j}+{p}_{j-1}\vert\,\vert\,\cF_\zs{j-1}\right),\label{sec.Tool.12}\\[3mm]
{Z}_\zs{4,j}=
\vert {Z}_\zs{1,j}\vert-
\E\,\left(\vert {Z}_\zs{1,j}\vert\,\vert\,\cF_\zs{j-1}\right)=\vert {Z}_\zs{1,j}\vert-\sqrt{2/\pi}.%\label{sec.Tool.13}
\end{cases}
\end{equation}
The sequences $( {Z}_\zs{3,j})$ and $( {Z}_\zs{4,j})$ will help to find the Dood decomposition of our approximation terms.
In order to represent the limit of transaction costs, we introduce 
\begin{equation}\label{sec:Bat.3-0}
\begin{cases}
G(a) =\E\,
\left( \left\vert Z+a\right\vert \right)
=2\varphi ( a) +a\left(2\Phi ( a) -1\right),\\[2mm]
\Lambda (a)=\E\,
\left( \vert Z+a\vert -\E\, \vert Z+a\vert \right)^2= 1+a^2-G^2(a),
\end{cases}
\end{equation}
for $a\in\bbr$ and $Z\sim \cN(0,1)$.
We also write $o( a_n^{-r})$ for generic sequences of random variables 
$(X_\zs{n})$ satisfying $
\P-\lim_\zs{n\rightarrow\infty }a_n^{r}\,X_\zs{n}=0$.

\subsection{Approximation for stochastic integrals}\label{Discr}
For any $L>0$, we consider the stopping time 

\begin{equation}\label{sec:Discr.1}
\tau^{*}=
\tau^{*}_\zs{L}=\inf
\left\{t\ge 0: \sigma(y_\zs{t})+\vert \sigma'(y_t)\vert>L
\right\},
\end{equation}
and denote by $S^{*}_\zs{t}=S_\zs{\tau^*\wedge t} $ and
$y^{*}_t=y_\zs{\tau^*\wedge t}$ the corresponding stopped processes. We provide an approximation procedure for It\^o stochastic integrals through the sequences $(Z_\zs{1,j})$ and $(Z_\zs{2,j})$. The discrete approximation concerns the class of functions satisfying the technical condition,

%\vspace{3mm}
\noindent $(\H)$ {\em $A: \bbr_\zs{+}\times\bbr_\zs{+}\times \bbr\to\bbr$ is a continuously differentiable
function satisfying the following: there exist $\gamma>0$ and a positive function
$U$ such that for any $x\ge 0, y\in\bbr$,
$$
\sup_\zs{\lambda>0}\min(\lambda^{\gamma},1)\vert A(\lambda, x,y)\vert\le U(x,y)
\quad\mbox{and}\quad \sup_{0\le t\le 1}\E\, (S^{*}_\zs{t})^{m}U^{2r}(S^{*}_\zs{t},y^{*}_\zs{t})<\infty,$$
for any $-\infty<m<+\infty$, $r\ge 0$ and $L>0$.
}
\begin{remark}
 We can check directly that for $k\ge 2$, $\partial_x^k\wh{C}(\lambda,x)= x^{k-1}\lambda^{-k/2} \wt{\varphi}(\lambda,x) P(\ln(x/K)),$
where $P$ is some polynomial. 
Therefore, all functions $A_0$ appearing in the approximation below are of the form $\lambda^{-k/2}x^m \bar{\sigma}(y) P(\ln(x/K)),$
where $\bar{\sigma}$ can be a power of ${\sigma}$ or of its two first derivatives $\sigma'$, $\sigma{''}$.
In bounded volatility settings, it can be shown with some computational effort (see e.g.,
\cite{Darses, Lepinette08, Lepinette10}) that 
\begin{equation}\label{sec:Discr.2}
\sup_\zs{0 \le t\le 1}\E S_t^m \ln^{2r}S_t<\infty, \quad \mbox{for any}\quad m\in \bbr, \,r\ge0. 
\end{equation}
The latter property is, however, not always fulfilled for SV models with unbounded volatility. In fact, it has been demonstrated in \cite{And-Pit, Lions-Mus} that the stock price does not admit integrable moments in general SV markets, unless some natural conditions are imposed on the correlation and the volatility dynamics coefficients. Thus, asymptotic analysis using $L^2$ estimates as in the existing works may be impossible in general SV frameworks. Nevertheless, note that \eqref{sec:Discr.2} is true for processes stopped by $\tau^{*}$. Below, the approximation analysis will be established in the sense of convergence in probability, in order to avoid this integrability issue.
\end{remark}
For simplicity, we use the notation $\check{S}=(S,y)$. The following technique is frequently applied in our asymptotic analysis.
\begin{proposition}\label{Pr.Tool.3} Let $A(\lambda,x,y)=A_0(\lambda,x,y)\wt{\varphi}(\lambda,x)$, where $A_0=A_0(\lambda,x,y)$ is a function satisfying $(\H)$. Then, for $i=1,2$,
 \begin{equation}\label{sec:Est.4}
  \int_\zs{0}^1\wh{\sigma}_\zs{t}^2
\left(\int_\zs{t}^1\, 
A(\lambda_\zs{t},\check{S}_\zs{u})\d W_\zs{u}^{(i)}\right) \d t
=\varrho^{-1}\sum_\zs{j=m_\zs{1}}^{m_2}\ov{A}_\zs{j-1}
\,{Z}_\zs{i,j}
\Delta \lambda_\zs{j}+o(\theta_\zs{n}^{-1}),
 \end{equation}
where $\theta_\zs{n}=n^{\beta}\varrho^{2\beta}$, $\ov{A}_\zs{j}=\ov{A}(\lambda_\zs{j},\check{S}_\zs{t_\zs{j}})$
and $\ov{A}(\lambda,x,y)=\int_{\lambda}^\infty A(z,x,y)\d z$.
\end{proposition}
\proof 
By making use of the stochastic Fubini theorem, we get
$$
\wh{I}_\zs{n}=
\int_\zs{0}^1\wh{\sigma}_\zs{t}^2
\left(\int_\zs{t}^1\, 
A(\lambda_\zs{t},\check{S}_\zs{u})\d W_\zs{u}^{(i)}\right) \d t
=
\int_\zs{0}^1
\left(
\int^{u}_\zs{0}\,
\wh{\sigma}_\zs{t}^2\,A(\lambda_\zs{t},\check{S}_\zs{u})\d t
\right)
\,\d W_\zs{u}^{(i)}\,.
$$
Then, changing the variables $v=\lambda_\zs{t}$ for the inner integral yields
$$\int^{u}_\zs{0}\,
\wh{\sigma}_\zs{t}^2\,A(\lambda_\zs{t},\check{S}_\zs{u})\d t=
 \int^{\lambda_\zs{0}}_\zs{\lambda_\zs{u}}\,A(v,\check{S}_\zs{u})\d v
=\ov{A}(\lambda_\zs{u},\check{S}_\zs{u})-\ov{A}(\lambda_\zs{0},\check{S}_\zs{u}).$$ In other words,
$
\wh{I}_\zs{n}=\wh{I}_\zs{1,n}
-
\wh{I}_\zs{2,n}$, where 
$\wh{I}_\zs{1,n}=\int_\zs{0}^1\,\check{A}_\zs{u}\,\d W_\zs{u}^{(i)}$, 
$\check{A}_\zs{u}=\ov{A}(\lambda_\zs{u},\check{S}_\zs{u})$
and $\wh{I}_\zs{2,n}
=\int_\zs{0}^1\,\ov{A}(\lambda_\zs{0},\check{S}_\zs{u})\,\d W_\zs{u}^{(i)}$. 
Moreover, we have
\begin{equation}\label{sec:Est.4-00}
\wh{I}_\zs{1,n}=\int_\zs{0}^{t^*} \check{A}_\zs{u} \d W_\zs{u}^{(i)}+
\int_\zs{t^*}^{t_\zs{*}} \check{A}_\zs{u} \d W_\zs{u}^{(i)}
+\int_\zs{t_\zs{*}}^{1} \check{A}_\zs{u} \d W_\zs{u}^{(i)}
:=R_\zs{1,n}+ R_\zs{2,n}+R_\zs{3,n}\,.
\end{equation}
Let $\e>0$ and $b>0$. One observes that $\P( \theta_\zs{n}\vert \wh{I}_\zs{2,n}\vert >\e)$ is bounded by
$\P(\tau^{*}_\zs{L}<1)+\P( \theta_\zs{n}\vert \wh{I}_\zs{2,n}\vert >\e, \, 
\tau^{*}_\zs{L}=1).
$
By condition $(\C_\zs{1})$, we have 
\begin{equation}\label{sec:Est.4-00-1}
 \limsup_\zs{L\to\infty} \P ( \tau^{*}_\zs{L}<1)=0\,.
\end{equation}
In view of $(\H)$, one deduces $\ov{A}(\lambda_\zs{0},x,y)|\le C \sqrt{K}\wt{U}(x,y)e^{-\lambda_\zs{0}/8}$,
% \begin{align*}
% |\ov{A}(\lambda_\zs{0},x,y)|\le C \sqrt{K x^{-1}}|U(x,y)|\int^{\infty}_\zs{\lambda_\zs{0}}e^{-\lambda/8}\d \lambda \le C \sqrt{K}\wt{U}(x,y)e^{-\lambda_\zs{0}/8},
% \end{align*}
where $\wt{U}(x,y)={x}^{-1/2}U(x,y)$.
Now, putting $\check{A}^{*}_\zs{u}=\check{A}_\zs{u\wedge\tau^{*}}$
 and $\wh{I}^{*}_\zs{2,n}=\int_\zs{0}^1\,\check{A}^{*}_\zs{u}\,\d W_\zs{u}^{(i)}$, one has
$\P( \theta_\zs{n}\vert \wh{I}_\zs{2,n}\vert >\e, \, \tau^{*}_\zs{L}=1)=
\P( \theta_\zs{n}\vert \wh{I}^{*}_\zs{2,n}\vert >\e)\,.
$
Using the Chebychev inequality, we obtain 
\begin{align*}
\P( \theta_\zs{n}\vert \wh{I}^{*}_\zs{2,n}\vert >\e)
\le \e^{-2}\theta_\zs{n}^2 \E\,(\wh{I}^{*}_\zs{2,n})^2\le  C \e^{-2}\theta_\zs{n}^2 e^{-\lambda_\zs{0}/8} 
\sup_{0\le t\le 1}\,
\E \,\wt{U}^2(\check{S}^{*}_\zs{t}).
\end{align*}
Hence, due to condition $(\H)$, 
$\wh{I}_\zs{2,n}=o(\theta_\zs{n}^{-1})$ as $n\to\infty$. Similarly, taking into account that 
$l^{*}\le \lambda_\zs{u}\le \lambda_\zs{0}$ for $0\le u\le t^{*}$,
we get $R_\zs{1,n}=o(\theta_\zs{n}^{-1})$. 

Next, let us show the same 
behavior for the last 
term in \eqref{sec:Est.4-00}. Indeed, for some fixed $\eta>0$ and $L>0$, one has
\begin{equation}\label{sec:Est.4-1}
 \P \left(\theta_\zs{n}\vert R_\zs{3,n}\vert >\varepsilon\right)
\le 
\P \left(\theta_\zs{n}\vert R_\zs{3,n}\vert >\varepsilon, \, 
\Gamma_\zs{1,\eta,L} \right)+\P\left(\Gamma^{c}_\zs{1,\eta,L}\right),
\end{equation}
where 
$ \Gamma_\zs{1,\eta,L}= 
\left\{\inf_{t_*\le u\le 1}\vert  \ln(S_u/K)\vert>\eta, \,\tau^{*}_\zs{L}=1\right\}$.
Then, by taking into account Lemma \ref{Le.Tool.3-0} and the integrability condition $(\C_1)$, one gets 
$${\lim}_\zs{\eta\to 0}\ov{\lim}_\zs{n\to\infty}\ov{\lim}_{ L\to\infty} \P(\Gamma^{c}_\zs{1,\eta,L})=0.$$
On $\Gamma_\zs{1,\eta,L}$, we have $\check{A}=\check{A}^{*}$
and 
\begin{align*}
&\vert\check{A}^{*}_\zs{u}\vert \le U(\check{S}^{*}_\zs{u})\int_\zs{\lambda_\zs{u}}^{\infty} (1+z^{-\gamma})\wt{\varphi}(z,S_\zs{u}^{*})  \d z
\le \wt{U}(\check{S}^{*}_\zs{u}) \check{f}^{*}_u,
\end{align*}
where $\check{f}^{*}_u=\sqrt{K/(2\pi)}\int_\zs{\lambda_\zs{u}}^{\infty} (1+z^{-\gamma}) e^{-{\eta^2}/(2z)-{z}/{8}} \d z$. Set 
$\Gamma_\zs{3,j}=\{\vert\check{A}_\zs{u}\vert \le \wt{U}(\check{S}^{*}_\zs{u}) \check{f}^{*}_u\},$
$ \wh{A}^{*}_\zs{u}=\check{A}^{*}_\zs{u} {\bf 1}_\zs{\Gamma_\zs{3,j}}$ and $\wh{R}_\zs{3,n}=\int_\zs{t_\zs{*}}^{1} \wh{A}^{*}_\zs{u} \d W_\zs{u}^{(i)}.
$
By Chebychev's inequality, we obtain
\begin{align*}
\P \left(\theta_\zs{n}\vert R_\zs{3,n}\vert >\varepsilon, \Gamma_\zs{1,\eta,L} \right)\le {\theta_\zs{n}^2}{\varepsilon ^{-2}}\int_\zs{t_\zs{*}}^{1} \E (\wh{A}^{*}_\zs{u})^2 \d u\le
{\theta_\zs{n}^2}{\varepsilon ^{-2}}\sup_\zs{0\le u\le 1}\E\,{\wt{U}^2(\check{S}_\zs{u}^{*})} 
\int_\zs{t_\zs{*}}^{1} (\check{f}^{*}_u)^2\d u,
\end{align*}
which converges to zero as 
$\int_\zs{t_\zs{*}}^{1} (\check{f}^{*}_u)^2\d u
\le C{\lambda_\zs{0}^{-4\beta}} l_\zs{*}.
$
Hence, $R_\zs{3,n}=o(\theta_n^{-1})$. It remains to discretize the integral term $R_\zs{2,n}$ via the sequence $(Z_{i,j})$. 
The key steps for this aim are the followings. First, we represent
$
R_\zs{2,n}=\int_{t^*}^{t_*} \check{A}_\zs{u}\d W_\zs{u}^{(i)}=
\sum_{j=m_\zs{1}}^{m_\zs{2}}\int_{t_{j-1}}^{t_j} 
\check{A}_\zs{u}\d W_\zs{u}^{(i)}
$
and replace the It\^o integral in the last sum with 
$\ov{A}_\zs{j-1} Z_{i,j}\sqrt{\Delta t_j}$.
%where $\ov{A}(\lambda,x,y)=\int_{\lambda}^\infty A(z,x,y) \d z$.
 Next, Lemma \ref{Le.Tool.1} enables us to substitute $\sqrt{\Delta t_j}={\varrho}^{-1}\Delta \lambda_j$ into the last sum to obtain the martingale ${\mathcal M}_{m_\zs{2}}$ defined by
$
{\mathcal M}_k={\varrho}^{-1}\sum_{j=m_\zs{1}}^{k}\ov{A}_{j-1} Z_{i,j}\Delta \lambda_j.
$
We need to show that
$
 \vert R_\zs{2,n} -{\mathcal M}_{m_\zs{2}}\vert=o(\theta_\zs{n}^{-1})	
$
or equivalently, $
\sum_{j=m_\zs{1}}^{m_\zs{2}} B_\zs{j,n}=o(\theta_n^{-1}),
$
where $B_\zs{j,n}=\int_{t_{j-1}}^{t_j}
\wt{A}_\zs{u,j}\d W_\zs{u}^{(i)}$ and $\wt{A}_\zs{u,j}=\bar{A}(\lambda_\zs{u},\check{S}_\zs{u})-\bar{A}(\lambda_\zs{j-1},\check{S}_\zs{t_\zs{j-1}}).$
For this aim, we introduce the set
$$
\Gamma_\zs{2,b}=\left\{\sup_\zs{t^{*}\le u\le 1} \sup_\zs{z\in \bbr} \left(\vert A(z,\check{S_\zs{u}})\vert + \left\vert {\partial}_\zs{x}  \bar{A}(z,\check{S_\zs{u}})\right\vert +\left\vert {\partial}_\zs{y}  \bar{A}(z,\check{S_\zs{u}})\right\vert\right)\le b\right\}.
$$
Then, for any $\varepsilon>0$, $\P\left( \theta_\zs{n}\vert \sum_{j=m_\zs{1}}^{m_\zs{2}} B_\zs{j,n}\vert > \varepsilon\right)$ is bounded by
$\P(\Gamma_\zs{2,b}^c)+\P(\tau^{*}<1)+\varpi_n,$ where 
$\varpi_n=\P\left( \theta_\zs{n}\vert \sum_{j=m_\zs{1}}^{m_\zs{2}} B_\zs{j,n}\vert > \varepsilon, \,\Gamma_\zs{2,b},\, \tau^{*}=1\right).
$
Let $\wh{B}_\zs{j,n}=\int_{t_{j-1}}^{t_j}
\wh{A}_\zs{u,j}\d W_\zs{u}^{(i)}$, where
$$\wh{A}_\zs{u,j}=\wt{A}_\zs{u,j}{\bf 1}_\zs{\{\vert \wt{A}_\zs{u,j}\vert\le b(\vert \lambda_u-\lambda_\zs{j-1}\vert+\vert{S}_\zs{u}^{*}-S_\zs{t_\zs{j-1}}^{*}\vert +\vert{y}_\zs{u}^{*}-y_\zs{t_\zs{j-1}}^{*}\vert)\}}.$$
Then, $
\varpi_n=\P\left( \theta_\zs{n}\vert \sum_{j=m_\zs{1}}^ {m_\zs{2}}\wh{B}_\zs{j,n}\vert>\varepsilon\right)$, which is smaller than
$\varepsilon^{-2}\theta_\zs{n}^2 \sum_{j=m_\zs{1}}^{m_\zs{2}}\E \, \wh{B}_\zs{j,n}^2
$
by Chebychev's inequality. Clearly, $\E \,\wh{B}_\zs{j,n}^2$ is bounded by
\begin{align*}
3 b^2\left(\int_{t_{j-1}}^{t_j}((\lambda_\zs{u}-\lambda_\zs{j-1})^2+
\E({S}_\zs{u}^{*}-S_\zs{t_\zs{j-1}}^{*})^2+\E({y}_\zs{u}^{*}-y_\zs{t_\zs{j-1}}^{*})^2)\d u\right)\le (\Delta \lambda_j)^3+ (\Delta t_j) ^2
\end{align*}
up to a multiple constant. Consequently, $\theta_\zs{n}^2 \sum_{j=m_\zs{1}}^{m_\zs{2}}\E \,\wh{B}_\zs{j,n}^2\le C\theta_\zs{n}^2 \sum_{j=m_\zs{1}}^{m_\zs{2}} (\Delta \lambda_j)^3+ (\Delta t_j)^2$, which converges to 0 by Lemma \ref{Le.Tool.1} and condition $(\C_\zs{2})$. On the other hand, by Lemma \ref{Le.Tool.3-2}, we get $\lim_\zs{b\to\infty}\, \ov{\lim}_\zs{n\to\infty} \P(\Gamma_\zs{2,b}^c)=0$. The proof is complete.\endproof

%\vspace{5mm}
%\noindent The following assertion guarantees that functions apprearing in the proofs of Main Results hold the technical conditions $(\H_0)$ and $(\H)$.
%\begin{lemma}\label{Le.Tool.4}
 %Let $\wh{C}(\lambda,x)$ be the solution to \eqref{sec:Rev.7}. Then for any $k\ge 2$, the $k$-th derivative of $\wh{C}(\lambda,x)$ satisfies condition $(\H)$ and condition $(\H_0)$ with respect to the first argument. The same conclusion holds for the set ${\cal D},$ where
%$$
%{\cal D}=\frac{1}{\wh{\sigma}^2(t)}\left\{\wh{C}_\zs{t},\wh{C}_{tx},\wh{C}_{txx},\dots\right\}.
%$$
%\end{lemma}
%\proof See the Appendix \ref{Event}.

\subsection{Limit theorem for approximations}\label{CTL}
 We first recall the following result in \cite{Hall}, which is useful for studying asymptotic distributions of discrete martingales. 

\begin{theorem} \label{Th.sec:Bat.0}[Theorem 3.2 and Corollary 3.1, p.58 in \cite{Hall}]
 Let ${\mathcal M}_\zs{n}=\sum_\zs{i=1}^{n}X_\zs{i}$ be a zero-mean, square
integrable martingale and $\varsigma $ be an a.s. finite random variable.
Assume that the following convergences are satisfied in probability: 
$$
\sum_\zs{i=1}^{n}\E\left( X_\zs{i}^{2}\boldsymbol{1}_\zs{\left\{ \left\vert
X_\zs{i}\right\vert >\delta \right\} }|\mathcal{F}_\zs{i-1}\right) \longrightarrow
0\quad\mbox{ for any}\quad\delta >0 \quad\mbox{and}\quad \sum_\zs{i=1}^{n}\E\left( X_\zs{i}^{2}|\mathcal{F}_\zs{i-1}\right)
\longrightarrow \varsigma^{2}.
$$
Then, $({\mathcal M}_\zs{n})$ converges in law to $X$ whose
characteristic function is $\E\exp ( -\frac{1}{2}\varsigma
^{2}t^{2}),$ i.e., $X$ has a Gaussian mixture distribution.
\end{theorem}

In this subsection, we establish special versions of Theorem \ref{Th.sec:Bat.0}. In fact, our aim is to study the asymptotic distribution of discrete martingales resulting from approximation \eqref{sec:Est.4} in Proposition \ref{Pr.Tool.3}. First,  we define 
\begin{equation}\label{sec:Bat.4}
{\mathcal M}_\zs{k}=\sum_\zs{j=m_\zs{1}}^k\upsilon_\zs{j}, \quad m_\zs{1}\le k\le m_\zs{2},
\end{equation}
where $\upsilon_\zs{j}=\sum^{3}_\zs{i=1}
{A}_\zs{i,j-1}
\,{Z}_\zs{i,j}
\Delta \lambda_\zs{j}$, 
${A}_\zs{i,j}={A}_\zs{i}(\lambda_\zs{j},\check{S}_\zs{t_\zs{j}})$
and $Z_{i,j}$ defined as in \eqref{sec:Tool.12-1} and in \eqref{sec.Tool.12}.
To describe the asymptotic variance of $(\cal M)$, we introduce the following function
\begin{align}\nonumber
\L( \lambda, x,y)&=
A^{2}_\zs{1}(\lambda,x,y)
+
2A_\zs{1}(\lambda,x,y) 
A_\zs{3}(\lambda,x,y) 
( 2\Phi({p}) -1) \\[2mm] \label{sec:Bat.5}
&+A^{2}_\zs{3}(\lambda, x,y)\,\Lambda({p})
%(1+{p}^{2}-G^{2}({p}))
+
A^{2}_\zs{2}(\lambda, x,y)\,,
\end{align}
where ${p}$ is defined in \eqref{sec:Bat.3-1}. Set 
\begin{equation}
\check{\mu}=\frac{1}{2}(\mu +1)\wt{\mu}^{\frac{2}{\mu +1}} \quad\mbox{and}\quad \wh{\mu}=(\mu -1)/(\mu +1).
\end{equation}

\begin{proposition}\label{Pr.sec:Bat.1}Let $A_{i}^{0}=A_{i}^{0}(\lambda,x,y), \, i=1,2,3$ be functions having property $(\H)$ and $A_\zs{i}(\lambda,x,y)=A_{i}^{0}(\lambda,x,y)\wt{\varphi}(\lambda,x)$. Then, for any fixed $\varrho>0$, 
the sequence  
$
(n^{\beta }{\mathcal M}_\zs{m_\zs{2}})_\zs{n\ge 1}
$
 weakly converges to a 
mixed Gaussian variable with mean zero and variance $\varsigma^{2}$ defined
as
$\varsigma^{2}=\varsigma^{2}(\check{S}_\zs{1})=\check{\mu} \varrho ^{\frac{2}{\mu +1}
}\int_\zs{0}^{+\infty }\lambda^{\wh{\mu}}\L(\lambda,\check{S}_\zs{1}) \d\lambda$.
The same property still holds if some (or all) of the functions $A_i$ are of the form $\int_{\lambda}^\infty A_{i}^{0}(z,x,y)\wt{\varphi}(z,x) \d z$.
\end{proposition}

\proof Note that the square integrability property is not guaranteed for $(\upsilon_\zs{j})$. To overcome this issue, we consider the ``stopped version'' 
$(\upsilon^{*}_\zs{j})$, which is obtained by substituting 
$\check{S}_\zs{t_{j-1}}$ by $\check{S}^{*}_\zs{t_{j-1}}$ in $A_\zs{i}$, i.e.,
%in $X_\zs{j}$, i.e. 
%$X_\zs{j}^{*}=\E\,({\upsilon_\zs{j}^{*}}^{2}\boldsymbol{1}
%_\zs{\left\{ \left\vert \upsilon_\zs{j}^{*}\right\vert >\delta\right\}} \vert {\cal F}_\zs{j-1})
%$, where
 $ \upsilon_\zs{j}^{*}=\sum^{3}_\zs{i=1}{A}_\zs{i}(\lambda_\zs{j},\check{S}^{*}_\zs{t_\zs{j}})\,{Z}_\zs{i,j}
\Delta \lambda_\zs{j}$. Let
${\mathcal M}_\zs{k}^{*}=\sum_\zs{j=m_\zs{1}}^k\upsilon_\zs{j}^{*}$, the corresponding 
stopped martingale. First, we show, throughout Theorem \ref{Th.sec:Bat.0}, that for any $L>0$,
this martingale weakly converges to
a mixed Gaussian variable with mean zero and variance 
$\varsigma^{*2}(L)=\varsigma^{2}(\check{S}^{*}_\zs{1})$. To this end,
 setting $ \Gamma_\zs{1,\eta}= 
\left\{\inf_{t^{*}\le u\le 1}\vert  \ln(S_u^{*}/K)\vert>\eta\right\}$
 and 
$
\a^{*}_\zs{j}=\E\,(\upsilon^{*2}_\zs{j}\,\boldsymbol{1}
_\zs{\{ \vert \upsilon^{*}_\zs{j}\vert >\delta\}} \vert {\cal F}_\zs{j-1}),
$ we obtain
\begin{align}\label{sec:Est.4-1-0}
\P\left(n^{2\beta} \vert\sum_\zs{j=m_\zs{1}}^{m_\zs{2}}
\a^{*}_\zs{j}\vert>\varepsilon\right)\le \P\left(n^{2\beta} \vert\sum_\zs{j=m_\zs{1}}^{m_\zs{2}}
 \a^{*}_\zs{j}\vert>\varepsilon,\,\Gamma_\zs{1,\eta}\right)+\P(\Gamma^{c}_\zs{1,\eta}).
\end{align}
It suffices to show that the first probability on the right side of \eqref{sec:Est.4-1-0} converges to zero. 
Indeed, from the proof of Proposition \ref{Pr.Tool.3}, 
one observes that on the set $\Gamma_\zs{1,\eta}$,
\begin{equation}\label{sec:Bat.5-1}
\max_{i=1,2,3}\left\vert A_i(\lambda_\zs{u}, \check{S}_\zs{u}^{*})\right\vert\le \wt{U}(\check{S}_\zs{u}^{*})(1+\lambda_\zs{u}^{-\gamma}), \quad t^{*}\le u\le t_\zs{*},
\end{equation} 
for some $\gamma>0$ and $\wt{U}(\check{S})=S^{-1/2}{U}(\check{S})$. 
Set $\wh{\upsilon}_\zs{j}^{*}=\upsilon_\zs{j}^{*}\boldsymbol{1}_\zs{\Gamma_\zs{3,j}}$ and $\wh{\a}^{*}_\zs{j}=\E\,(\wh{\upsilon}_\zs{j}^{*2}\boldsymbol{1}_\zs{\{ \vert \wh{\upsilon}_\zs{j}^{*}\vert >\delta\}} 
\vert {\cal F}_\zs{j-1})$, where
$$
\Gamma_\zs{3,j}=\left\{\max_\zs{1\le i\le 3}\left\vert 
A_i(\lambda_\zs{u}, \check{S}_\zs{u}^{*})\right\vert
\le \wt{U}(\check{S}^{*}_u)(1+\lambda_\zs{u}^{-\gamma})\right\}.$$
We have
$$
\P\left(n^{2\beta} \vert\sum_\zs{j=m_\zs{1}}^{m_\zs{2}}
 \a^{*}_\zs{j}\vert>\varepsilon,\,\Gamma_\zs{1,\eta,L}\right)=\P\left(n^{2\beta} \vert\sum_\zs{j=m_\zs{1}}^{m_\zs{2}}
 \wh{\a}^{*}_\zs{j}\vert>\varepsilon\right)\le \varepsilon^{-1}n^{2\beta} \sum_\zs{j=m_\zs{1}}^{m_\zs{2}}
\E\, \wh{\a}^{*}_\zs{j},
$$
by Markov's inequality. By using Chebychev's inequality and then again Markov's inequality, we observe that 
$$
\E\, \wh{\a}^{*}_\zs{j}\le \sqrt{\E\,\wh{\upsilon}_\zs{j}^{*4}}\sqrt{\P(\vert \wh{\upsilon}_\zs{j}^{*}\vert >\delta)} \le \delta^{-2}\E\,\wh{\upsilon}_\zs{j}^{*4} 
\le 9\delta^{-2} (1+\lambda_\zs{u}^{-\gamma})^4(\Delta \lambda_\zs{j})^4 \E \,\wt{U}^4(\check{S}^{*}_u)\sum_{i=1}^3
\,{Z}^4_\zs{i,j}.
$$
Note that $Z_\zs{i,j}$ has bounded moments. Then, by using \eqref{sec:Bat.5-1}, 
we obtain $\varepsilon^{-1}\,n^{2\beta}\, \sum_\zs{j=m_\zs{1}}^{m_\zs{2}}
\E\, \wh{\a}^{*}_\zs{j}$ is bounded by
$
9\varepsilon^{-1}\delta^{-2} n^{2\beta}\,
 \sum_\zs{j=m_\zs{1}}^{m_\zs{2}}
(1+\lambda_\zs{u}^{-\gamma})^4  (\Delta \lambda_\zs{j})^4,
$
which converges to zero by Lemma \ref{Le.Tool.1}.

We now verify the limit of the sum of conditional variances $\E(\upsilon_\zs{j}^{*2}|\mathcal{F}_\zs{j-1})$. Set $\upsilon_\zs{i,j}^{*}={A}^{*}_\zs{i,j-1}
\,{Z}_\zs{i,j}\,\Delta \lambda_\zs{j}$. As ${Z}_\zs{1,j}$ and ${Z}_\zs{2,j}$ are independent,
$
\E\left(\upsilon_\zs{1,j}^{*}\upsilon_\zs{3,j}^{*}\vert\mathcal{F}_\zs{j-1}\right) =
\E\left( \upsilon_\zs{2,j}^{*}\upsilon_\zs{3,j}^{*}|\mathcal{F}_\zs{j-1}\right)
=0.
$
It follows that
\begin{align*}
\E( \upsilon_\zs{j}^{*2}|\mathcal{F}_\zs{j-1})=\E( \upsilon_\zs{1,j}^{*2}|\mathcal{F}_\zs{j-1})
+\E( \upsilon_\zs{2,j}^{*2}|\mathcal{F}_\zs{j-1})
+\E( \upsilon_\zs{3,j}^{*2}|\mathcal{F}_\zs{j-1})
+2\E( \upsilon_\zs{1,j}^{*}\upsilon_\zs{2,j}^{*}|\mathcal{F}
_\zs{j-1}) .
\end{align*}
Now, observe that for $Z\sim N( 0,1) $ and some constant $a$, $
\E( Z\left\vert Z+a\right\vert) =2\Phi (a)
-1$ and $
\E\,\left( Z+a\right)^{2} -
\left(\E\vert Z+a\right\vert)^{2}=\Lambda(a).
$
On the other hand, $\Delta {
\lambda}_\zs{j}=n^{-2\beta}(
1+o( 1) ) \check{\mu}\,\varrho ^{\frac{2}{\mu +1}}\lambda_\zs{j-1}^{\wh{\mu }}$ by Lemma \ref{Le.Tool.1}. Therefore, 
$$
n^{2\beta }\E( \upsilon_\zs{j}^{*2}|\mathcal{F}_\zs{j-1})=(
1+o( 1) ) \check{\mu}\,\varrho ^{\frac{2}{\mu +1}}\,
\lambda_\zs{j-1}^{\wh{\mu }}\,\L(\lambda_\zs{j-1},\check{S}_\zs{t_\zs{j-1}}^{*}) \Delta {
\lambda}_\zs{j}.
$$
By Lemma \ref{Le:Bat.5},
$n^{2\beta}\,\sum_\zs{j=m_\zs{1}}^{m_\zs{2}}
\E( \upsilon_\zs{j}^{*2}|\mathcal{F}_\zs{j-1})$
converges in probability to $\varsigma ^{*2}(L)$.
% defined as 
%$\check{\mu} \varrho ^{\frac{2}{\mu +1}}\int_\zs{0}^{+\infty }
%\lambda^{\wh{\mu}}\L(\lambda,\check{S}_\zs{1}^{*}) \d\lambda.$
Thus, $n^{\beta}{\mathcal M}_\zs{m_2}^{*}$ weakly converges to ${\cal N}(0,\varsigma ^{*2}(L))$ by Theorem~\ref{Th.sec:Bat.0}. 
Moreover, property  \eqref{sec:Est.4-00-1} implies that for any $\delta>0$,
$$
\lim_\zs{L\to\infty}\ov{\lim_\zs{n\to\infty}}\P(n^{\beta}\vert {\mathcal M}_\zs{m_2}-{\mathcal M}_\zs{m_2}^{*}\vert>\delta)=0\,.
$$
Therefore, by taking into account that $\varsigma ^{*2}(L)$ converges a.s. to $\varsigma^{2}$ as $L\to\infty$, we conclude that $n^{\beta} {\mathcal M}_\zs{m_2}$ converges in law to ${\cal N}(0,\varsigma^2)$. This completes the proof.\endproof

\vspace{1mm}

Next, we study the asymptotic property of the following martingale 
\begin{equation}
\ov{{\mathcal M}}_k=\sum_\zs{j=m_\zs{1}}^{k}\left({A}_\zs{1,j-1}
\,{Z}_\zs{1,j}+{A}_\zs{2,j-1}\,{Z}_\zs{2,j}
+{A}_\zs{4,j-1}
\,{Z}_\zs{4,j}\right)
\Delta \lambda_\zs{j}. 
\end{equation}
The limiting variance will be defined throughout the function
\begin{align}
\ov{\L}( \lambda, x,y)=
A^{2}_\zs{1}(\lambda,x,y)+A^{2}_\zs{2}(\lambda, x,y)
+(1-2/\pi)A^{2}_\zs{4}(\lambda, x,y).
\end{align}
The following result is similar to Proposition \ref{Pr.sec:Bat.1}.
\begin{proposition}\label{Pr.sec:Bat.3} Let $A_{i}^{0}=A_{i}^{0}(\lambda,x,y), \, i=1,2,4$ be functions having property $(\H)$ and $A_\zs{i}(\lambda,x,y)=A_{i}^{0}(\lambda,x,y)\wt{\varphi}(\lambda,x)$. Then, for any fixed $\varrho>0$ 
the sequence  
$
(n^{\beta }\,\ov{{\mathcal M}}_\zs{m_\zs{2}})_\zs{n\ge 1}
$
weakly converges to a 
mixed Gaussian variable with mean zero and variance $\ov{\varsigma}^{2}$ given by
$
\ov{\varsigma}^{2}=\check{\mu}\, \varrho ^{\frac{2}{\mu +1}
}\int_\zs{0}^{+\infty }\lambda^{\wh{\mu}}\,\ov{\L}(\lambda,\check{S}_\zs{1}) \d\lambda.
$
The same property still holds if some (or all) $A_i$ are of the form $\int_{\lambda}^\infty A_{i}^{0}(z,x,y)\wt{\varphi}(z,x) \d z$.
\end{proposition}

\proof The conclusion follows directly from the proof of Proposition \ref{Pr.sec:Bat.1} and the observation that 
$\E\, {Z}_\zs{4,j}^2=\E(\vert {Z}_\zs{1,j}\vert-\sqrt{2/\pi})^2=1-{2}/{\pi},$ and 
$ \E \,({Z}_\zs{i,j}{Z}_\zs{4,j})=0,
$ for $i=1,2$ and $m_\zs{1}\le j\le m_\zs{2}$.\endproof

\vspace{1mm}
In the rest of the subsection, we establish a limit theorem for a martingale of the following form
$$
\check{{\mathcal M}}_\zs{k}=\sum_\zs{j=m_\zs{1}}^k\left({A}_\zs{1,j-1}
\,{Z}_\zs{1,j}+{A}_\zs{3,j-1}
\,{Z}_\zs{3,j}\right)\,
\Delta \lambda_\zs{j}:=\sum_\zs{j=m_\zs{1}}^k\check{\upsilon}_\zs{j}, \quad m_1\le k \le m_2,
$$
 where  $A_\zs{i}(\lambda,x,y)=A_{i}^{0}(\lambda,x,y)\wt{\varphi}(\lambda,x)$ and $A_{i}^{0}, \, i=1,3$ are functions having property $(\H)$. The following result is helpful for the case when $\varrho$ diverges to infinity as in Theorem \ref{Th.sec:Mar.2}.
%Let us consider 
%\begin{equation}\label{sec:Bat.6}
%\check{{\mathcal M}}_\zs{k}=\sum_\zs{j=m_\zs{1}}^k\left({A}_\zs{1,j-1}
%\,{Z}_\zs{1,j}+{A}_\zs{3,j-1}
%\,{Z}_\zs{3,j}\right)\,
%\Delta \lambda_\zs{j}:=\sum_\zs{j=m_\zs{1}}^k\check{\upsilon}_\zs{j},\quad  m_\zs{1}\le k\le m_\zs{2},
%\end{equation}
%where functions $A_\zs{1}$ and $A_\zs{3}$ are independent of the third argument.

\begin{proposition}\label{Pr.sec:Bat.2}Under condition $(\C_\zs{2})$, the sequence $
\left(n^{\beta}\,\varrho^{\frac{-1}{\mu+1}}\,\check{{\mathcal M}}_\zs{m_\zs{2}}\right)
$
weakly converges to a mixed Gaussian variable with mean zero and variance 
$
\check{\varsigma}^2=\check{\mu}\int_\zs{0}^{+\infty }\lambda ^{\wh{\mu}}
\,{\check{\L}}(\lambda,S_\zs{1} ) \d\lambda ,
$
where
${\check{\L}}(\lambda,x,y) =A^{2}_\zs{1}(
\lambda, x,y)
+2A_\zs{1}(\lambda,x,y) A_\zs{3}( \lambda, x,y) +
A^{2}_\zs{3}(\lambda,x,y)$.
The same property still holds if some (or all) $A_i$ are of the form $\int_{\lambda}^\infty A_{i}^{0}(z,x,y)\wt{\varphi}(z,x) \d z$.
\end{proposition}

\proof
We determine the limit of conditional variances of $n^{\beta}\,\varrho^{\frac{-1}{\mu+1}}\,\check{{\mathcal M}}_\zs{m_\zs{2}}$. We first observe that
\begin{equation}\label{sec:Bat.7}
n^{2\beta}\varrho^{\frac{-2}{\mu+1}}\,\E(\check{\upsilon}_\zs{j}^{2}|\mathcal{F}_\zs{j-1}) =
\check{\mu} ( 1+o( 1) )\, \lambda_\zs{j-1}^{\wh{\mu}}\,{\check{\Q}} (\lambda
_\zs{j-1},\check{S}_\zs{t_\zs{j-1}}) \Delta \lambda_\zs{j},
 \end{equation}
where
$
{\check{\Q}}(\lambda,x,y)=
A^{2}_\zs{1}(\lambda,x,y)+A^{2}_\zs{3}(\lambda,x,y)
\,\Lambda ({p})
+2A_\zs{1}(\lambda,x,y)A_\zs{3}(\lambda,x,y)\left(
2\Phi ( \left\vert {p}\right\vert )-1\right).$
Moreover, it can be checked directly that the function $G(\cdot)$ defined in
\eqref{sec:Bat.3-0} satisfies the following inequalities: $|a|\le G(a)\le |a|+2\varphi(a)\,$, for any $a\in\bbr$.
This implies that
$
\left\vert \Lambda (a)-1\right\vert
\le 4 
|a|\varphi(a)+\varphi^{2}(a),
$
hence,
$\sup_\zs{a\in\bbr}|\Lambda (a)|<\infty$. Note also that ${\check{\Q}}\to {\check{\L}}$ a.s. as $n\to\infty$ because ${p}(\lambda,x,y)\to\infty$ as $\varrho =\varrho (n)\to\infty$, for any $x>0$ and $\lambda\neq 2\ln (x/K)$. Using Lemma~\ref{Le:Bat.5}, we claim that 
the sum of the terms on the right-hand side of \eqref{sec:Bat.7} converges in probability to $\check{\varsigma}^{2}$. The proof is completed by running again the argument in the proof of Proposition 
\ref{Pr.sec:Bat.1}.\endproof

\subsection{Proof of Theorem \ref{Th.sec:Mar.1}}
We first observe that $I_\zs{1,n}$ approaches $2\min (S_\zs{1},K)$ at order $\theta_\zs{n}$. In particular, 
setting
$\bar{I}_\zs{1,n}=\int_\zs{0}^{1}{\lambda_\zs{t}}^{-1/2}\wh{\sigma}_\zs{t}^{2}
\,\left( 
S_\zs{t}\wt{\varphi}(\lambda_\zs{t},S_\zs{t}) -S_\zs{1}\wt{
\varphi}(\lambda_\zs{t},S_\zs{1})
\right)
\, 
\d t$
and changing variables $v=\int_t^1\wh{\sigma}^2_s\d s$,
we can represent $
I_\zs{1,n}=S_\zs{1}
\int_\zs{0}^{\lambda_\zs{0}}{v}^{-1/2}\wt{\varphi}(v,S_\zs{1})\, \d v
+
\bar{I}_\zs{1,n}+o(\theta_n^{-1})
\,.
$
The first integral in the right side converges a.s. to $2\min(S_1,K)$ by \eqref{sec:Rev.4-4}, while $\bar{I}_\zs{1,n}$ is approximated by $\int_\zs{0}^{1}\wh{\sigma}_\zs{t}^{2}
\,\left( 
\int_t^1 \sigma(y_\zs{u})S_\zs{u}H(\lambda_\zs{t},S_\zs{u}) \d W_\zs{u}^{(1)}
\right)
\, 
\d t,$ where $H=(2^{-1}\lambda^{-1/2}-\lambda^{-3/2}\ln(x/K))\wt{\varphi}( \lambda,x)$.
The discretization technique of Proposition \ref{Pr.Tool.3} can be applied to replace the latter double integral by ${\cal U}_\zs{1,m_2}$, defined by
\begin{equation}\label{Sec:I1I2.1-1}
{\cal U}_\zs{1,k}=\varrho^{-1}\sum_\zs{j=m_1}^k\sigma ( y_\zs{t
_\zs{{j-1}}}) S_\zs{t_\zs{j-1}}\,
\check{H}_\zs{j-1}\,
 Z_\zs{1,j}\,\Delta 
{\lambda}_\zs{j},\quad m_1\le k\le m_2,
\end{equation}
where $\check{H}(\lambda,x)=\int_\lambda^\infty(z^{-1/2}/2-z^{-3/2}\ln(x/K))\wt{\varphi}(z,x) \d z.$
We summarize the asymptotic form of $I_\zs{1,n}$ in the following.
\begin{proposition}\label{Le.sec:I1I2.1}
If $\varrho $ either is constant or satisfies condition $(\C_\zs{2}) $ then,
$$
\P-\lim_\zs{n\longrightarrow \infty }\theta_\zs{n}
\left\vert I_\zs{1,n}-2\min (S_\zs{1},K)
-
{\cal U}_\zs{1,m_2}
\right\vert
=0.
$$
\end{proposition}

\vspace{2mm}
\noindent Next, we claim that $I_\zs{2,n}=o(\theta_\zs{n}^{-1})$.
\begin{proposition}\label{Pr.sec:I1I2.2}
If $\varrho $ either is a positive constant or satisfies condition $(\C_\zs{2}) $, then $\theta_\zs{n}I_\zs{2,n}$ converges to zero in probability as $n\to\infty$.
\end{proposition}
\proof See the Appendix \ref{sec:Appendix.3.2}.\endproof

\vspace{1mm}

Let us study the trading volume $J_n$. First, it is easy to check that for any $v>0,\ 1-\Phi (v)\le v^{-1}\varphi(v)$. Now, observe that
$
 \vert \gamma^n_\zs{t_i}-\gamma^n_\zs{t_{i-1}}\vert\le  \vert 1-\gamma^n_\zs{t_i}\vert+\vert 1-\gamma^n_\zs{t_{i-1}}\vert,
$
which almost surely converges to zero more rapidly than any power of $n$ when $\inf_\zs{1\le i\le n} \lambda_\zs{i}\ge l^*\Longleftrightarrow i\le m_1$. The same property can be deduced for the case $\sup_\zs{i}\lambda\le l_*\Longleftrightarrow i\ge m_2$. To see this, we note that for $\lambda_u\le l_*$, $S_u(\omega)=S_\zs{1-(\lambda_u/\lambda_0)^{4\beta}}(\omega)$ converges to $S_1(\omega)$ as $n\to\infty$ uniformly in $\lambda_u\in[0,l_*]$, for any $\omega$ outside the zero probability set $\{S_1=K\}$. Therefore, one can truncate the sum and keep only the part corresponding to index $m_1\le j\le m_2$. 
In other words, $J_n$ is approximated by $J_\zs{1,n}=\sum_\zs{j=m_1}^{m_2}S_\zs{t_\zs{j}}\left\vert \Delta\Phi_\zs{j}\right\vert.$ 
Putting $ b_\zs{j}=\left|\Delta\Phi_\zs{j}  \right\vert -
\wt{\varphi}_\zs{j-1} \left\vert\Delta \v_\zs{j}\right\vert$,
 we can represent $J_\zs{1,n}=J'_\zs{1,n}+\varepsilon_\zs{1,n}+\varepsilon_\zs{2,n}\,,$
where $J'_\zs{1,n} =\sum_\zs{j=m_1}^{m_2}S_\zs{t_\zs{j-1}}\wt{\varphi}_\zs{j-1}
\left\vert\Delta\v_\zs{j}\right\vert$, $\varepsilon_\zs{1,n} =\sum_\zs{j=m_1}^{m_2}
\Delta S_\zs{t_\zs{j-1}} \left\vert 
\Delta_\zs{j} \Phi\right\vert$ and $\varepsilon_\zs{2,n}=\sum_\zs{j=m_1}^{m_2}\,S_\zs{t_\zs{j-1}}\,b_\zs{j}$.
% $$
% J'_\zs{1,n} =\sum_\zs{j=m_1}^{m_2}S_\zs{t_\zs{j-1}}\wt{\varphi}_\zs{j-1}
% \left\vert\Delta\v_\zs{j}\right\vert,\quad\varepsilon_\zs{1,n} =\sum_\zs{j=m_1}^{m_2}
% \Delta S_\zs{t_\zs{j-1}} \left\vert 
% \Delta_\zs{j} \Phi\right\vert,\quad \varepsilon_\zs{2,n}=\sum_\zs{j=m_1}^{m_2}\,S_\zs{t_\zs{j-1}}\,b_\zs{j}.
% $$ 
In view of \eqref{sec:Est.1-2} and 
condition $(\C_\zs{2}$), $\varepsilon_\zs{1,n}=o(\theta_n^{-1})$ as $n\to\infty$. 
Furthermore, by using the Taylor expansion, we obtain 
$
\left\vert \varepsilon_\zs{2,n}\right\vert \leq 
S_\zs{\sup}
\sum_\zs{j=m_1}^{m_2}
\,\left\vert \Delta\v_\zs{j}\right\vert ^{2}
$
up to a multiple constant, where $S_\zs{\sup}=\sup_\zs{0\le t\le 1}S_\zs{t}$. Now, by taking into account that
$$
\E \left\vert \v_\zs{j-1}-\v_\zs{j}\right\vert ^{2}\le 
\frac{1}{n\lambda_\zs{j-1}}
+\left(\lambda_\zs{j-1}^{1/2}- \lambda_\zs{j}^{1/2}\right)^{2}
+\left(\lambda_\zs{j-1}^{-1/2}- \lambda_\zs{j}^{-1/2}\right)^{2}
$$
up to a multiple constant and using condition $(\C_\zs{2}$) together with \eqref{sec:Est.1-2},
we get
$\vert\varepsilon_\zs{2,n}\vert=o(\theta_\zs{n}^{-1}).$
Next, by using It\^o's Lemma and the substitution ${\lambda}_j=\lambda_\zs{0}(1-t_j)^{4\beta}$, we replace $J'_\zs{1,n}$ by 
\begin{equation}\label{Sec:I1I2.1-2}
J_\zs{2,n}=\sum_\zs{j=m_1}^{m_2}{\lambda}_\zs{j-1}^{-1/2}S_\zs{t_\zs{j-1}}\wt{\varphi}_\zs{j-1}
 \vert 
{\varkappa}_\zs{j}\vert \Delta {\lambda}_\zs{j}:=\sum_\zs{j=m_1}^{m_2}{\zeta}_\zs{j},\quad{\varkappa}_\zs{j}=\varrho^{-1}
\sigma(y_\zs{t_\zs{j-1}}) Z_\zs{1,j}+q_\zs{j-1},
\end{equation}
where $q$ is defined in \eqref{sec:Rev.4-2-1}.
We need to determine the limit of $J_n$ throughout the Doob's decomposition of $J_\zs{2,n}$ w.r.t. the filtration 
$\left(\mathcal{F}_\zs{j}\right)_\zs{m_1\leq j\leq m_2}$.
To this end,
note that
\begin{equation*}\label{sec:Trc.4}
\E({\zeta}_\zs{j}\vert \mathcal{F}_\zs{j-1})
={\lambda}_\zs{j-1}^{-1/2}S_\zs{t_\zs{j-1}}\wt{\varphi}_\zs{j-1} \Delta {\lambda}_\zs{j}
\,\E(\vert {\varkappa}_\zs{j}\vert|
\mathcal{F}_\zs{j-1}),
\end{equation*}
\noindent 
where 
$\E(\left\vert {\varkappa}_\zs{j}\right\vert |\mathcal{F}
_\zs{j-1})=\varrho^{-1}\sigma(y_\zs{t_\zs{j-1}})G({p}_\zs{j-1})
:=D_\zs{j-1}
$
%\begin{equation*}\label{sec:Trc.5}
%\E(\left\vert {\varkappa}_\zs{j}\right\vert |\mathcal{F}
%_\zs{j-1})=\varrho^{-1}\sigma(y_\zs{t_\zs{j-1}})G({p}_\zs{j-1})
%:=\G({\lambda}_\zs{j-1},S_\zs{t_\zs{j-1}},y_\zs{t_\zs{j-1}})\,,
%\end{equation*}
and $G({p})$ defined in \eqref{sec:Bat.3-0}. 
Let
\begin{equation}\label{sec:Trc.4-0}
B(\lambda,x,y)={\lambda}^{-1/2}x\wt{\varphi}({\lambda}, x)D({\lambda},x,y) \quad \mbox{and}
\quad J_\zs{3,n}=\sum_\zs{j=m_1}^{m_2}B_\zs{j-1}\Delta {\lambda}_\zs{j}.
\end{equation}
 We observe that
$J_\zs{2,n}=J_\zs{3,n}+{\cal U}_\zs{2,m_2}\,,$
where 
\begin{equation}\label{sec:Trc.6}
{\cal U}_\zs{2,k}=\sum_\zs{j=m_1}^{k}
{\lambda}_\zs{j-1}^{-1/2}
S_\zs{t_\zs{j-1}}\wt{\varphi}
_\zs{j-1}\ov{\varkappa}_\zs{j}\Delta {\lambda}_\zs{j}\quad\mbox{and}\quad \ov{\varkappa}_\zs{j}:=\left\vert{\varkappa}_\zs{j}\right\vert
-D_\zs{j-1}.
\end{equation}
By substituting $\check{S}_\zs{t_{j-1}}$ by $\check{S}_1$ everywhere in ${J}_\zs{3,n}$, we write ${J}_\zs{3,n}={J}_\zs{4,n}+{J}_\zs{5,n}$, where 
${J}_\zs{4,n}=\sum_\zs{j=m_1}^{m_2}B(\lambda_\zs{j-1},\check{S}_\zs{1})\Delta {\lambda}_\zs{j}$,
${J}_\zs{5,n}=\sum_\zs{j=m_1}^{m_2}B_\zs{j-1}^{*}\Delta {\lambda}_\zs{j}$
and $B_\zs{j-1}^{*}= B(\lambda_\zs{j-1},\check{S}_\zs{t_{j-1}})-B(\lambda_\zs{j-1},\check{S}_\zs{1})$. 
Then, by Lemma \ref{Le.sec:Bat.1}, we can check that ${J}_\zs{4,n}$ converges a.s. to $J({S}_\zs{1},{y}_\zs{1},\varrho)$ at rate $\theta_\zs{n}$. 
Now, an application of It\^o's Lemma for $B_\zs{j-1}^{*}$ yields stochastic integrals with respect to the Wiener processes. Owing to Proposition \ref{Pr.Tool.3}, the sum of these integrals can be approximated by ${\cal U}_\zs{3,m_2}$, defined by
$$
{\cal U}_\zs{3,k}={\varrho }^{-1}\sum_{i=1}^2\sum_\zs{j=m_1}^{k}Q_\zs{i,j-1} Z_\zs{i,j}\Delta {\lambda}_\zs{j}, \quad m_1\le k\le m_2,
$$
where $Q_1=\int_\lambda^\infty (x\sigma(y)\partial_x B +\r F_2(t(\lambda),y)\partial_y B)\d z$ and $Q_2=\sqrt{1-{\r} ^2}  
F_2 (t(\lambda),y)\int_\lambda^\infty  \partial_y B\d z$ and $t(\lambda)=1-(\lambda/\lambda_\zs{0})^{4\beta}$. The asymptotic form of $J_{n}$ is summarized in the following. 
\begin{proposition}
\label{Pr.sec:Trc.2}
For any fixed $\varrho>0$, 
$$
\P-\lim_\zs{n\longrightarrow \infty }\theta_\zs{n}
\left\vert J_\zs{n}-J({S}_\zs{1},{y}_\zs{1},\varrho)-({\cal U}_\zs{2,m_2}+{\cal U}_\zs{3,m_2})\right\vert=0.
$$

\end{proposition}

\noindent Now, the martingale part of the
hedging error is given by ${\cal M}_\zs{m_2}$, defined by
$$
{\cal M}_\zs{k}=\frac{1}{2}{\cal U}_\zs{1,k}-\kappa_\zs{*}({\cal U}_\zs{2,k}+{\cal U}_\zs{3,k})
={\varrho }^{-1}
\sum_\zs{j=m_1}^{k}\sum_{i=1}^3{A}_\zs{i,j-1}Z_\zs{i,j}\Delta {\lambda}_\zs{j}, \quad m_1\le k\le m_2,
$$
where $A_\zs{1} =-\sigma ( y)\,x\check{H}/2$, $A_\zs{2}=\kappa_\zs{*}{Q}_\zs{2}$ and
${A}_\zs{3} =-\kappa_\zs{*}\sigma(y)\lambda ^{-1/2}x\wt{
\varphi}(\lambda,x)$. It is easy to see that the assumption of Proposition~\ref{Pr.sec:Bat.1} is fulfilled for $A_i,\,i=1,2,3$. Hence, the sequence $\left(n^{\beta }{\cal M}_\zs{m_2}\right)_\zs{n\ge 1}$
converges in law to a mixed Gaussian variable by Proposition~\ref{Pr.sec:Bat.1}, which proves Theorem~\ref{Th.sec:Mar.1}. \endproof

%\subsection{Proof of Theorem~\ref{Th.sec:Mar.1}}

\subsection{Proof of Theorem~\ref{Th.sec:Mar.2}}
When $\varrho\to\infty$ under condition $(\C_\zs{2})$, the approximation for $J_{n}$ is slightly different. In particular, observe first that for any $b\in\bbr$, $\E\left\vert
aZ+b\right\vert $ can be approximated by $b(2\Phi(b/a)-1) $ as $a\to 0$. 
%enough small\footnote{In fact, 
%$\E\left\vert aZ+b\right\vert \to\vert b\vert $ as $ a\to 0$ 
%but this may leads us to using the generalized It\^o's formula since $\vert x \vert$ 
%is not a $C^2$-function.}
Therefore, we can replace $J_\zs{3,n}$ in \eqref{sec:Trc.4-0} by the sum
$
\wh{J}_\zs{3,n}=\sum_\zs{j=m_1}^{m_2}\,\wh{B}_\zs{j-1} \Delta {\lambda}_\zs{j},$
where 
%\begin{equation*}\label{sec:Trc.7}
$\wh{B}(\lambda,x)
=\lambda^{-1/2}\,x\,\wt{\varphi}(\lambda,x)q(\lambda,x)
\wt{\Phi}(\varrho q(\lambda,x)),$
%\end{equation*} 
with $\wt{\Phi}(q)=2\Phi(\varrho\,q)-1$ and $q(\lambda,x)$ defined in \eqref{sec:Rev.4-2-1}. 
Puting
$\wh{J}_\zs{4,n}=\sum_\zs{j=m_1}^{m_2}\wh{B}({\lambda}_\zs{j-1},S_\zs{1})\, 
\Delta {\lambda}_\zs{j}$ and %\begin{equation*}\label{sec:Trc.8}
$\wh{B}_\zs{j-1}^*= \wh{B}({\lambda}_\zs{j-1},S_\zs{t_\zs{j-1}})
-\wh{B}({\lambda}_\zs{j-1},S_\zs{1})$,
%\end{equation*}
we represent $\wh{J}_\zs{5,n}:=\wh{J}_\zs{3,n}-\wh{J}_\zs{4,n}=\sum_\zs{j=m_1}^{m_2}\wh{B}_\zs{j-1}^*\Delta {\lambda}_\zs{j}$.
Now, using Lemma~\ref{Le.sec:Bat.1}, we can directly show that
$\vert\wh{J}_\zs{4,n}-J^{*}(S_\zs{1})\vert=o(\theta_\zs{n}^{-1})$.
\noindent Furthermore, owing to It\^o's formula, we replace
$\wh{B}_\zs{j-1}^*$ by $\int_{t_\zs{j-1}}^1\partial_x\wh{B}
({\lambda}_\zs{j-1},S_\zs{u})\d S_\zs{u}.$ Direct calculations yield that
\begin{align*}
{\partial_x\wh{B}}=\lambda^{-1/2}\,
\wt{\varphi}(\lambda,x)[ -2q^2(\lambda,x)\wt{\Phi}(\lambda,x)
+\frac{1}{2\lambda}\wt{\Phi}(\lambda,x)+\frac{\varrho}{\lambda}\varphi(\varrho\,
q(\lambda,x))].
\end{align*}
\noindent
Clearly, $\wt{\Phi}(\varrho q)\to \sign(q)$ and $\varphi(\varrho\,
q) \to 0$ as $\varrho\to\infty$. Now, using Proposition \ref{Pr.Tool.3}, we can approximate $\wh{J}_\zs{5,n}$ by $\wh{\cal U}_\zs{3,m_\zs{2}}$, defined by
$ \wh{\cal U}_\zs{3,k}=
\varrho^{-1}\sum_\zs{j=m_1}^{k}
\,
\sigma ( y_\zs{t_\zs{j-1}}) S_\zs{t_\zs{j-1}}N_\zs{j-1}\,Z_\zs{1,j}\Delta\lambda_\zs{j},
$
where
$N(\lambda,x)
=\int_\zs{\lambda}^{+\infty }z^{-1/2}{\wt{\varphi}(z,x)}
 \left(-2q^2(z,x)+1/(2z)\right)\sign(q(z,x))\d z$.
The asymptotic representation of the trading volume is summarized in the following.
\begin{proposition}\label{Pr.sec:Trc.1}Under conditions $(\C_1)-(\C_2)$,
$$
\P-\lim_\zs{n\longrightarrow \infty }\theta_\zs{n}\vert J_\zs{n}-J^{\ast }(S_\zs{1}) -({\cal U}_\zs{2,m_2}+\wh{\cal U}_\zs{3,m_2}
)\vert=0.
$$
\end{proposition}
\noindent Now, the martingale part $\varrho^{-1}\check{{\cal M}}_\zs{m_2}$ of the hedging error is determined by 
$$
\check{{\cal M}}_\zs{k}
=\frac{{\varrho}}{2}{\cal U}_\zs{1,k}-\kappa_\zs{*}{\varrho}({\cal U}_\zs{2,k}+\wh{\cal U}_\zs{3,k})
=\sum_\zs{j=m_1}^{k}(\check{A}_\zs{1,j-1}Z_\zs{1,j}+\check{A}_{3,j-1} Z_\zs{3,j})
\Delta {\lambda}_\zs{j} \label{thej-}, 
$$ where $\check{A}_i,\,i=1,2$ are explicitly determined and satisfy the assumption of Proposition \ref{Pr.sec:Bat.2}. Then, in view of $
\theta_\zs{n}{
\varrho }^{-1}\check{{\cal M}}_\zs{m_2}=n^{\beta }\varrho ^{-\frac{1}{\mu +1}}\check{{\cal M}}_\zs{m_2},$ Theorem \ref{Th.sec:Mar.2} is proved throughout Proposition \ref{Pr.sec:Bat.2}.
\endproof

\subsection{Proof of Theorem~\ref{Th.sec:Lepi.1}}
The key technique in Proposition~\ref{Pr.Tool.3} can be used to obtain a smart martingale approximation for the sum 
$\sum_{i\ge 1}\Delta S_\zs{t_{i}}\int_\zs{0}^{t_{i-1}}\wh{C}_\zs{xt}(u,S_\zs{u})\d u.
$
\begin{proposition}\label{Pr.sec:I1I2.3}
If $\varrho $ either is a positive constant or satisfies condition $(\C_\zs{2})$, then
$
\vert\ov{I}_\zs{2,n}-\ov{{\cal U}}_{1,m_2}\vert=o(\theta_\zs{n}^{-1}),
$
where $Y(\lambda,x)=\int_\lambda^\infty z^{-3/2}\ln(x/K)\wt{\varphi}(z,x)\d z$ and 
$$
\ov{{\cal U}}_\zs{1,k}=\varrho^{-1}\sum_\zs{j=m_1}^k\sigma ( y_\zs{t
_\zs{{j-1}}}) S_\zs{t_\zs{j-1}}\,
Y_\zs{j-1}\,
 Z_\zs{1,j}\,\Delta 
{\lambda}_\zs{j},\quad m_1\le k\le m_2.
$$
\end{proposition}
\proof The proof follows from the substitution $\Delta S_\zs{t_j}$ by $\varrho^{-1}\sigma(y_\zs{t_{j-1}}) S_\zs{t_{j-1}} \Delta \lambda_\zs{t_j}$ as in Proposition \ref{Pr.Tool.3}.\endproof

Let us now study the trading volume $\ov{J}_\zs{n} $ by following the procedure in the approximation of $J_n$. First, by It\^o's lemma,
$$
\ov{\gamma}_\zs{t_i}-\ov{\gamma}_\zs{t_{i-1}}=\int_{t_{i-1}}^{t_i}\wh{C}_{xx}(u,S_\zs{u})\d S_\zs{u}+\frac{1}{2}\int_\zs{t_{i-1}}^{t_i}\wh{C}_{xxx}(u,S_\zs{u})\sigma^2(y_\zs{u})S^2_\zs{u} \d u,
$$
where the time-correction, which involves the term $q_{j-1}$ in the formula of ${\varkappa}_j$ defined by \eqref{Sec:I1I2.1-2}, has been removed.
We now approximate $\ov{J}_n$ by 
%\begin{equation}\label{sec:Trc.2-0}
$$
\ov{J}_\zs{1,n}=\varrho^{-1}\sum_\zs{j=m_1}^{m_2}
\ov{B}_\zs{j-1}\left\vert Z_\zs{1,j} \right\vert \Delta {\lambda}_\zs{j}
\quad\mbox{and} \quad\ov{B}(\lambda,x,y)=\sigma (y)\,x\,{\lambda}^{-1/2}\wt{\varphi}(\lambda,x).
$$
%\end{equation}
As $\E\vert Z\vert =\sqrt{2/\pi}$, for $Z\sim {\cal N}(0,1)$, the Dood' decomposition of $\ov{J}_\zs{1,n}$ is given by 
$\ov{J}_\zs{2,n}+ \bar{\cal U}_\zs{2,m_\zs{2}}$, where 
$
\ov{J}_\zs{2,n}=\varrho^{-1}\sqrt{2/\pi}\sum_\zs{j=m_1}^{m_2}\ov{B}_\zs{j-1} \Delta {\lambda}_\zs{j}$
and $\bar{\cal U}_\zs{2,m_\zs{2}}=\varrho^{-1}\sum_\zs{j=m_1}^{m_2}\ov{B}_\zs{j-1}
 Z_{4,j} \Delta {\lambda}_\zs{j}.
$
Now, putting $\ov{B}_\zs{j-1}^{*}=
\ov{B}({\lambda}_\zs{j-1},\check{S}_\zs{t_\zs{j-1}})-\ov{B}({\lambda}_\zs{j-1},\check{S}_1)$, we write $\ov{J}_\zs{2,n}=\ov{J}_\zs{4,n}+\ov{J}_\zs{3,n}$,
where 
$$
\ov{J}_\zs{4,n}=\varrho^{-1}\sqrt{2/\pi}\sum_\zs{j=m_1}^{m_2}\ov{B}_\zs{j-1} \Delta {\lambda}_\zs{j},
\quad 
\ov{J}_\zs{3,n}=\varrho^{-1}\sqrt{2/\pi}\sum_\zs{j=m_1}^{m_2}\ov{B}_\zs{j-1}^{*}\Delta {\lambda}_\zs{j}.
$$
Observe that $\ov{J}_\zs{4,n}$ converges a.s. to $\eta\min(S_1,K)$ by Lemma \ref{Le.sec:Bat.1} and \eqref{sec:Rev.4-4}. 
We now find the suitable martingale approximation for $\ov{J}_\zs{3,n}$.
By It\^o's formula, $\ov{B}_\zs{j-1}^{*}$ can be replaced by 
$
\sum_{i=1}^{2}\int_t^1 \ov{Q}_\zs{i}({\lambda}_\zs{j-1},\check{S}_\zs{u}) \d W_u^{(i)}$,
where
$\ov{Q}_\zs{1}=\sigma(y)x\partial_x \ov{B}+\r F_2(t(\lambda),y)\partial_y\ov{B}$ and
$\ov{Q}_\zs{2}=\sqrt{1-{\r}^2}F_2(t(\lambda),y)\partial_y\ov{B}.
$ Direct calculations show that $
\partial_x \ov{B}=\sigma(y)(2^{-1}\lambda^{-1/2}-\lambda^{-3/2}\ln(X/K))\wt{\varphi}(\lambda,x)$ and 
$
\partial_y \ov{B}=\sigma'(y)\lambda^{-1/2}x\wt{\varphi}(\lambda,x).
$
Now, Proposition \ref{Pr.Tool.3} can be applied to approximate $\ov{J}_\zs{3,n}$ by the martingale $\ov{ \cal U}_{3,m_2}$, defined by 
$$\ov{ \cal U}_{3,k}=\varrho^{-1}\sum_\zs{j=m_1}^{k} (\ov{A}_\zs{1,j-1}Z_{1,j}+\ov{A}_\zs{2,j-1}Z_{2,j})
 \Delta {\lambda}_\zs{j}, \quad m_1\le k\le m_2,
$$
for explicit functions $\ov{A}_\zs{i}, i=1,2$.
The final asymptotic form of $\ov{J}_\zs{n} $ is given below.
\begin{proposition}\label{Pr.sec:Lepi.1}
 If $\varrho$ is a positive constant independent of $n$ then, 
$$
\P-\lim_\zs{n\rightarrow \infty }\theta_\zs{n}\vert\ov{J}_\zs{n}-\eta\min(S_1,K)-(\ov{\cal U}_\zs{2,m_2}+\ov{\cal U}_\zs{3,m_2})\vert=0.$$
\end{proposition}
\noindent Hence, the martingale part of the
hedging error for L\'epinette's strategy is determined by $\ov{{\cal M}}_\zs{m_\zs{2}}={\cal U}_\zs{1,m_\zs{2}}+\ov{{\cal U}}_\zs{1,m_\zs{2}}-\kappa_\zs{*}(\ov{ \cal U}_{2,m_\zs{2}}+\ov{ \cal U}_{3,m_\zs{2}})$. The latter martingale sum can be represented in the form
$$
\ov{{\cal M}}_\zs{k}={\varrho }^{-1}
\sum_\zs{j=m_1}^{k}({A}_\zs{1,j-1}Z_\zs{1,j}+
{A}_\zs{4,j}Z_\zs{4,j-1}+{A}_\zs{2,j-1}Z_\zs{2,j})\Delta {\lambda}_\zs{j},\quad m_1\le k\le m_2,
$$
for explicit functions ${A}_\zs{i}$ holding the assumption of Proposition~\ref{Pr.sec:Bat.3}. Then, $\left(n^{\beta }\ov{{\cal M}}_\zs{m_2}\right)_\zs{n\ge 1}$ converges in law to a mixed Gaussian variable, which completes the proof. \endproof

\section{Conclusion}
We studied the problem of approximate option replication in SV settings using a new specification for adjusted volatility. Although our model employed a simpler adjusted volatility than in the previous literature, we obtain the same asymptotic results for both Leland' and L\'epinette's strategies in general SV markets. A possible connection to high frequency markets with proportional transaction costs was also discussed. As an application, we showed that the option price inclusive of transaction costs can be reduced by adapting the theory of quantile hedging. Note that our approach is still helpful for more general settings, for example, when the friction rule admits a separate-variable representation \cite{Nguyen1}. This generalization includes the case where trading costs are based on the physical number of traded shares.
Lastly, in a companion paper, we extended the method to multidimensional frameworks for European options with general payoffs written on several assets \cite{Nguyen2}.

\vspace{3mm}
\noindent{\bf Acknowledgment}: The authors would like to thank the two referees and editor for remarks and suggestions that have helped to improve the paper. The first author wishes to express his gratitude to the Vietnam Overseas Scholarship Program
(project 322) for financial support.

\section*{Appendix}\label{sec:A}

\setcounter{section}{0}
\renewcommand{\thesection}{\Alph{section}}
\section{Auxiliary Lemmas}\label{sec:Appendix.1}
\begin{lemma}\label{Le.Tool.1} There exist two positive constants $C_1,C_2$ such that
\begin{equation}\label{sec:Est.1-2}
 C_1\,{n}^{-2\beta}\varrho^{\frac{2}{\mu+1}}\,\nu_0(l_*)\le \inf_\zs{m_\zs{1}\le j\le m_\zs{2}}\vert\Delta\lambda_j\vert\le
\sup_\zs{m_\zs{1}\le j\le m_\zs{2}}\vert\Delta\lambda_j\vert\le C_2{n}^{-2\beta}\varrho^{\frac{2}{\mu+1}}\,\nu_0(l^*),
 \end{equation} where $\nu_0(x)=x^{(\mu-1)/(\mu+1)}$. Moreover, for any $m_\zs{1}\le j\le m_\zs{2}$,
 \begin{equation}\label{sec:Est.1-1}
  \Delta\lambda_j={n}^{-2\beta}\varrho^{\frac{2}{\mu+1}}\,\nu_0(\lambda_{j-1})(1+o(1))\quad\mbox{and}\quad
{\Delta\lambda_j}\,(\Delta t_j)^{-1/2}=\varrho (1+o(1)).
\end{equation}
\end{lemma}
\proof It follows directly from the relation \eqref{sec:Bat.3}.\endproof

% One also deduces from Lemma \ref{Le.Tool.1} that
% \begin{equation}\label{sec:Est.1-2}
% \max_\zs{m_\zs{1}\leq j\leq m_\zs{2}}\vert\Delta \lambda_\zs{j}\vert\leq C n^{-2\beta }\varrho^{\frac{2}{\mu +1}}(l^{*}) ^{\frac{\mu-1}{\mu +1}}.
% \end{equation}
\noindent{\bf A technical condition $(\H_0)$}: {\em $A:\bbr_\zs{+}\to \bbr$ is a continuously differentiable 
function having absolutely integrable derivative $A'$  and
$$
\lim_\zs{n\to\infty} \theta_\zs{n}\,
\left( \int_\zs{0}^{l_\zs{\ast}}|A(\lambda)|\d \lambda
+\int_\zs{l^{*}}^{+\infty } |A(\lambda)|\d \lambda\right)=0,\quad \mbox{where}\quad \theta_\zs{n}=n^{\beta }\varrho ^{2\beta }.
$$
}
The following result is straightforward to check.
\begin{lemma}\label{Le.sec:Bat.1}
Let $\varrho$ either be a positive constant or satisfy condition $(\C_\zs{2})$. Then,
for any function $A$ satisfying condition $(\H_0)$,
\begin{equation}\label{Le.sec:Bat.1-0}
  \lim_\zs{n\to\infty}\theta_\zs{n}
\left\vert
\sum_\zs{j=m_\zs{1}}^{m_\zs{2}}{\bf 1}_\zs{\{\zs{\lambda_{j-1}}\ge a\}}
A(\lambda_\zs{j-1})\Delta\lambda_\zs{j}
-\int_a^{\infty}A(\lambda)\d\lambda
\right\vert =0.
\end{equation}
In particular, $ \lim_\zs{n\to\infty}\theta_\zs{n}
\left\vert
\sum_\zs{j=m_\zs{1}}^{m_\zs{2}}
A(\lambda_\zs{j-1})\Delta\lambda_\zs{j}
-\int_0^{\infty}A(\lambda)\d\lambda
\right\vert =0.$
\end{lemma}

\begin{lemma}\label{Le.Tool.3-0}For any $K>0$, $\lim_\zs{\varepsilon\to0}\limsup_\zs{v\to 1} \P(\inf_\zs{v\le u\le 1} \vert \ln(S_u/K)\vert \le \varepsilon )=0.$
% \begin{equation*}\label{sec:A.1.0-1}
% 
% \end{equation*}
\end{lemma}
\proof It follows from the fact that conditioning on the $\sigma$-field generated by the volatility process, the log-price process $\ln S_\zs{t}$ has Gaussian distribution.\endproof
% Let $\eta$ be some positive number. Clearly, $\P(\inf_\zs{v\le u\le 1} \vert \ln(S_u/K)\vert \le \varepsilon )$ is bounded by
% $$
% \P(\inf_\zs{v\le u\le 1} \vert \ln(S_u/K)\vert \le \varepsilon , \vert \ln(S_1/K)\vert > \eta)+\P( \vert \ln(S_1/K)\vert \le \eta).
% $$
% Let us show that the first probability is equal to zero for $v$ sufficiently close to $1$. Indeed, denoting $ \psi_u=\int_\zs{u}^{1}\sigma ^{2}(
% y_\zs{s}) \d s-\frac{1}{2}\int_\zs{u}^{1}\sigma ( y_\zs{s})\d W_\zs{s}^{(1)}$
% we can check that $\psi_v^*\to 0$ a.s. as $v\to 1$, where $\psi_v^*=\sup_\zs{v\leq u\leq 1}\left\vert \psi_u\right\vert$.
% So, if $\vert \ln(S_1/K)\vert > \eta$ then for $v\le u\le 1$,
% \begin{align*}
% \left\vert \ln (S_\zs{u}/K) \right\vert& =\left\vert \ln (
% S_\zs{1}/K) -\psi_u\right\vert \geq \left\vert \left\vert \ln (S_\zs{1}/K) \right\vert
% -\psi_v^* \right\vert 
% \geq \frac{1}{2}\left\vert \ln (
% S_\zs{1}/K) \right\vert>\eta/2.
% \end{align*}
% Therefore, for $\eta>2\varepsilon$, one obtains $\inf_\zs{v\le u\le 1} \vert \ln(S_u/K)\vert\ge \eta/2>\varepsilon$ and so, $$\P(\inf_\zs{v\le u\le 1} \vert \ln(S_u/K)\vert \le \varepsilon , \vert \ln(S_1/K)\vert > \eta)=0.$$
% Letting now $\eta\to0$ we get $\P( \vert \ln(S_1/K)\vert \le \eta)\to \P(S_1=K).$
% Note that conditioning on $\sigma$-field generated by the Wiener process driving $y$, 
% the log-price process $\ln S_\zs{t}$ has Gaussian distribution. Hence, $\P(S_1=K)=0$ and the proof is completed.\endproof
\begin{lemma}\label{Le.Tool.3-2}Suppose that $A_0$ and its derivatives 
$\partial_xA_\zs{0}, \partial_yA_\zs{0}$ verify condition $(\H)$. Set $A(\lambda,x,y)=A_0(\lambda,x,y)\wt{\varphi}(\lambda,x)$, $\bar{A}(\lambda,x,y)=\int_\zs{\lambda}^\infty A(z,x,y)\d z$ and define 
$$
r_\zs{n}=\sup_\zs{(z,r,d)\in [l_\zs{*}, l^{*}]\times{\cal B}}
\left(\vert \partial_\zs{\lambda}\bar{A}(z,r,d)\vert +\vert \partial_\zs{x}\bar{A}(z,r,d)\vert+\vert\partial_\zs{y}{A}(z,r,d)\vert\right)
,
$$ where
${\cal B}=[S_\zs{\min},S_\zs{\max}]\times [y_\zs{\min},y_\zs{\max}]$ with
$S_\zs{\min}=\inf_\zs{t^*\le u\le t_*}S_\zs{u},\, S_\zs{\max}=\sup_\zs{t^*\le u\le t_*}S_\zs{u}
$ 
and
$
y_\zs{\min}=\inf_\zs{t^*\le u\le t_*}y_\zs{u},\,
y_\zs{\max}=\sup_\zs{t^*\le u\le t_*}y_\zs{u}.
$
Then, $\lim_\zs{b\to \infty} \ov{\lim}_\zs{n\to\infty} \P( r_n >b)=0.$
\end{lemma}
\proof
Let $\varepsilon>0$. On the set $\Gamma_\zs{1,\varepsilon}=\{\inf_\zs{t^{*}\le u\le 1}\vert \ln(S_u/K)\vert \ge\varepsilon \}$, $$\sup_\zs{S_\zs{\min}\le r\le S_\zs{\max}}\wt{\varphi}(q,r)\le (2\pi)^{-1/2}\sqrt{Kr^{-1}} \exp\{-\varepsilon^2/(2q)-q/8\}.$$ By condition $(\H)$, there exists $\gamma>0$ such that 
\begin{equation*}
  \vert \bar{A}_\zs{x}(z,r,d)\vert\le C \vert \wt{U}(r,d)\vert\int_{z}^\infty (q^{-1/2}+q^{\gamma}) e^{-\varepsilon^2/(2q)-q/8} \d q \le C_\epsilon \wt{U}(r,d),
\end{equation*}
where $\wt{U}$ is some function verifying $\sup_\zs{0\le t\le 1}\E \,\wt{U}(\check{S_\zs{t}^{*}})<\infty$.
For any $\eta>0$ and $N>0$, let 
$$\Gamma_\zs{2,\eta}=\{\sup_\zs{(r,d)\in{\cal B}}\vert \wt{U}(r,d)-\wt{U}(\check{S}_\zs{1})\vert < \eta \}\bigcap \{ \vert \wt{U}(\check{S}_\zs{1})\vert < N\}.$$ It is clear that $\vert \wt{U}(r,d)\vert< N+\eta$ on the set $\Gamma_\zs{2,\eta}$. Similarly, taking into account $\partial_\zs{\lambda}\bar{A}(z,r,d)=-{A}(z,r,d),$ $\partial_\zs{y}\bar{A}(z,r,d)= \int_\zs{\lambda}^\infty \partial_\zs{y}A_0(z,x,y)\wt{\varphi}(z,x)$  we deduce that both $\vert\partial_\zs{\lambda}\bar{A}(z,r,d)\vert$ and $\vert\partial_\zs{y}\bar{A}(z,r,d)\vert$ are bounded on $\Gamma_\zs{2,\eta}$ by a constant $ C_\zs{N,\eta}$ independent of $b$.
Now, for $b>N+\eta+2C_\zs{N,\eta}$, $\P( r_n >b)$ is bounded by
$$
\P(\Gamma_\zs{1,\varepsilon}^c)+ \P(\sup_\zs{(r,d)\in{\cal B}}\vert \wt{U}(r,d)-\wt{U}(\check{S}_\zs{1}^{*})\vert \ge \eta)
+\P(\vert \wt{U}(\check{S}_\zs{1}^{*})\vert> N)+\P(\tau^{*}<1).
$$
By Lemma \ref{Le.Tool.3-0}, $\lim_\zs{\varepsilon\to0}\lim_\zs{n\to\infty}\P(\Gamma_\zs{1,\varepsilon}^c)\to 0$. Thanks to the continuity of the functions $S_t$ and $y_\zs{t}$, one gets $ \lim_\zs{n\to\infty}\P\left(\sup_\zs{(r,d)\in{\cal B}}\vert \wt{U}(r,d)-\wt{U}(\check{S}_\zs{1}^{*})\vert \ge \eta\right )=0$. Moreover, the integrability of $\wt{U}(\check{S}_\zs{1}^{*})$ implies that $\P(\vert \wt{U}(\check{S}_\zs{1}^{*})\vert> N)$ converges to zero as $N\to\infty$. By \eqref{sec:Est.4-00-1}, $\P(\tau^{*}<1)$ converges to 0 as $L\to\infty$, which completes the proof.\endproof

%\section{Proof of convergence lemmas}

\begin{lemma}\label{Le:Bat.5}
Let $\ov{A}(\lambda,x,y)=\int_\lambda^\infty A^{0}(z,x,y)\wt{\varphi}(z,x) \d z, \,\wt{A}=\ov{A}^2,$ where $A^{0}$ is a function having property $(\H)$. Then, for any $\gamma>0$,
$$\P-\lim_\zs{n\to\infty}\left\vert\sum_{j=m_1}^{m_2}\lambda_\zs{j-1}^\gamma\wt{A}(\lambda_\zs{j-1}, \check{S}_\zs{t_\zs{j-1}})
\Delta \lambda_j-\int_0^\infty \lambda^\gamma\wt{A}(\lambda, \check{S}_\zs{1}) \d\lambda\right\vert=0,
$$ where $\check{S}_\zs{t}=({S}_\zs{t},y_\zs{t})$. The same property still holds if $\ov{A}(\lambda,x,y)=A^{0}(\lambda,x,y)\wt{\varphi}(x,y)$ or is a product of these above kinds.
\end{lemma}
\proof
We prove for the first case $\ov{A}(\lambda,x,y)=\int_\lambda ^\infty A^{0}(z,x,y)\wt{\varphi}(z,x) \d z$, as the same argument can be made for the other cases. First, we split the expression under the absolute sign as $\sum_{j=m_1}^{m_2}\lambda_\zs{j-1}^\gamma\wt{A}(\lambda_\zs{j-1}, \check{S}_\zs{1})\Delta \lambda_j+ \sum_{j=m_1}^{m_2}\Delta_\zs{j,n}\Delta \lambda_j,$ where 
$
\Delta_\zs{j,n}=\wh{A}(\lambda_\zs{j-1}, \check{S}_\zs{t_\zs{j-1}})-\wh{A}(\lambda_\zs{j-1}, \check{S}_\zs{1})$ and $\wh{A}(\lambda,x,y)=\lambda^{\gamma}\wt{A}(\lambda,x,y).
$
It is clear that for any $(x,y)$, the function $\wh{A}(\cdot,x,y)$ satisfies condition $(\H_0)$. Hence, $\sum_{j=m_1}^{m_2}\wh{A}(\lambda_\zs{j-1}, \check{S}_\zs{1})\Delta \lambda_j$ converges a.s. to zero by Lemma \ref{Le.sec:Bat.1}. It remains to show that $\P( \vert\Delta_n\vert >\varepsilon)\to 0$ for any given $\varepsilon>0$, but it can be done by the same way as in Lemma \ref{Le.Tool.3-0}.\endproof 
\section{Proof of Proposition \ref{Pr.sec:I1I2.2}}\label{sec:Appendix.3.2}
The singularity of $\wh{C}$ at the maturity $T=1$ requires a separate treatment. Let $\varepsilon_\zs{n}=n^{-2\beta }\varrho ^{-4\beta }l_\zs{\ast}$. We then represent 
$I_\zs{2,n}=\int_\zs{0}^{1-\varepsilon_\zs{n}}\varpi_\zs{n}(t)\d W_\zs{t}^{(1)}+\int_\zs{1-\varepsilon_\zs{n}}^{1}\varpi_\zs{n}(t)\d W_\zs{t}^{(1)},
$ where 
$\varpi_\zs{n}(t)= ( \gamma_\zs{t}^{n}-
\wh{C}_\zs{x}(t,S_\zs{t})) \sigma (y_\zs{t}) S_\zs{t}$.
Taking into account that $\vert \gamma_\zs{t}^{n}-\wh{C}_\zs{x}( t,S_\zs{t})
\vert \leq 1,$ we obtain $\lim_\zs{n\to\infty}\, 
\theta_\zs{n}^2\,
\E\,\int_\zs{1-\varepsilon_\zs{n}}^{1}\varpi^{2}_\zs{n}(t)\d t=0\,.
$
Now put $\wh{t}_\zs{j}=\min ( t_\zs{j},1-\varepsilon_\zs{n})$. It then remains to prove that $\sum_\zs{j=1}^{n}\int_\zs{\wh{t}_\zs{j-1}}^{\wh{t}_\zs{j}}\E( \gamma
_\zs{t}^{n}-\wh{C}_\zs{x}( t,S_\zs{t})) ^{2}\d t=o(\theta_\zs{n}^{-2})$. 
% \begin{equation*}\label{sec:A.2}
% \lim_\zs{n\longrightarrow \infty }\theta_\zs{n}^2\sum_\zs{j=1}^{n}\int_\zs{\wh{t}_\zs{j-1}}^{\wh{t}_\zs{j}}\E( \gamma
% _\zs{t}^{n}-\wh{C}_\zs{x}( t,S_\zs{t})) ^{2}\d t=0,\quad \wh{t}_\zs{j}=\min ( t_\zs{j},1-\varepsilon_\zs{n}).
% \end{equation*}
Let us introduce the discrete sums
$w_\zs{1}( t) =\sum_\zs{j=1}^{n}{\lambda_\zs{t}}^{-1}( x_\zs{t}-x_\zs{
\wh{t}_\zs{j-1}}) ^{2} \xi_\zs{j}(t)$,  $w_\zs{2}( t) =\sum_\zs{j=1}^{n}\,x^{2}_\zs{t}\,
 (
\lambda_\zs{t}^{-1/2}-\lambda_\zs{\wh{t}_\zs{j-1}}^{-1/2})^{2}\,\xi_\zs{j}(t)
$ and $
w_\zs{3}( t) =\sum_\zs{j=1}^{n}( \lambda
_\zs{t}^{1/2}-\lambda_\zs{\wh{t}_\zs{j-1}}^{1/2})^{2}\,\xi_\zs{j}(t)$, where $\xi_\zs{j}(t)=\mathbf{1}_\zs{(\wh{t}_\zs{j-1}, \wh{t}
_\zs{j}]}(t)$ and $x_\zs{t}=\ln (S_\zs{t}/K)$.  Clearly,
%\begin{align*}
$\vert \gamma_\zs{t}^{n}-\wh{C}_\zs{x}( t,S_\zs{t})\vert
^{2}\leq  w_\zs{1}( t) +w_\zs{2}( t)
+w_\zs{3}(t).$ By taking into account that
$$
\sup_\zs{n,\,1\le j\le n}
\,n\,
\sup_\zs{0\le t\le 1}
\E( x_\zs{t}-x_\zs{\wh{t}_\zs{j-1}})^{2}\,\xi_\zs{j}(t)
<\infty
\quad \mbox{and}\quad\sup_\zs{0\le t\le 1}\,\E\,x^{2}_\zs{t}
<\infty,
$$  we have
$\theta_\zs{n}^2\E\int_\zs{0}^{1-\varepsilon_\zs{n}}w_\zs{1}( t) \d t 
\leq C n^{2\beta -3/2}\varrho ^{4\beta -1}$, which converges to zero by $(\C_\zs{2})$.
Now, the particular choice of $\varepsilon$ ensures that $\theta_\zs{n}^2
\E\int_\zs{0}^{1-\varepsilon_\zs{n}}w_\zs{2}(t) \d t \leq 
C{\theta_\zs{n}^2}n^{-2}(\varepsilon_\zs{n})^{-(4\beta+1)/4\beta}\lambda_\zs{0}^{-1}\,
,$ which tends to zero. The convergence for $w_3(t)$ can be shown in the same way.\endproof

\section{Moments of Orstein-Uhlenbeck's processes} \label{OU}
\begin{lemma}\label{Le.sec:A.4}
Suppose that $\sigma(z)\le \gamma(1+\vert z\vert)$ for all $z\in\bbr$, for some constant $\gamma>0$. Let $y_\zs{t}$ be an Orstein-Uhlenbeck process defined by
$
\d y_\zs{t}=(a-by_\zs{t})\d t+\d Z_\zs{t}
$
with some constants $a$ and $b>0$. Put $ X_\zs{\alpha}=\exp\left\{2\alpha \gamma^2\int_\zs{0}^1 y^2_s \d s\right\}$ and $\alpha_\zs{*}= {b^2}{(2\gamma^2(2b+a^2))}^{-1}$. Then, $\E X_\zs{\alpha}<\infty $ for any $0<\alpha<\alpha_\zs{*}$.
\end{lemma}
\proof Remark that $(a-by)y\le {a^2}/(2b)-by^2/2.
$
Then, by adapting Proposition 1.1.5 in \cite[p.24]{Kab-Ser}, we can show that 
$
\E\, \vert y_\zs{t}\vert^{2m}\le m! \left({2}/{b}+{a^2}/{b^2}\right)^m$ for any integer $ m\ge 1.
$
It follows that for any $0<\alpha<\alpha_\zs{*}$,
\begin{equation*}
\E X_\zs{\alpha}\le \sum_{m=0}{(\alpha 2\gamma^2)^m}({m!})^{-1}\E \,\vert y_\zs{t}\vert^{2m}\le 
\sum_{m=0}\left({2}/{b}+{a^2}/{b^2}\right)^m{(\alpha 2 \gamma^2)^m}<\infty.
\end{equation*}
If $y_\zs{t}$ is mean-reverting then $b$ takes very large values. Hence, it is possible to choose $\alpha>3/2+\sqrt{2}$ as discussed in Remark \ref{Re.sec:App.2}. \endproof


\begin{thebibliography}{99}
\bibitem{Ahn} Ahn H., Dayal M., Grannan E., Swindle G. (1998): Option
replication with transaction costs: General diffusion limits, {\it Ann.
Appl. Prob.}, {\bf8}(3), 767-707 .


\bibitem{And-Pit}Andersen L., Piterbarge V. (2007): Moment explosions in stochastic volatility models, {\it Finance and Stochastics}, {\bf11}, 29-50.

\bibitem{Bal-Rom}Ball C., Roma A. (1994): Stochastic volatility option pricing, {\it Finance and Quantitative Analysis}, {\bf 29}(4), 589-607.

\bibitem{BlackScholes}Black F., Scholes M. (1973): The pricing of
options and corporate liabilities, {\it J. Political Economy}, {\bf81}, 637-659.

\bibitem{Baran} Baran M. (2003): Quantile hedging on markets with
proportional transaction costs, {\it Appl. Math. Warsaw}, {\bf 30}, 193-208.

\bibitem{Barski}Barski M. (2011): Quantile hedging for multiple assets
derivatives, preprint available at \emph{http://arxiv.org/abs/1010.5810}.

\bibitem{Bratyk}Bratyk M., Mishura Y. (2008): The generalization of the
quantile hedging problem for price process model involving finite number of
Brownian and fractional Brownian motions, {\it Th. Stoch. Proc.}, {\bf 14}(3-4), 27-38.

\bibitem{Cvitanic}Cvitani\'{c} J., Karatzas I. (1996): Hedging and
portfolio optimization under transaction costs: a martingale approach,
{\it Mathematical Finance}, {\bf6}, 133-165.

\bibitem{Darses} Darses. S., L\'epinette L. (2011): Limit theorem for a modified Leland hedging strategy under constant transaction costs rate, {\it The Musiela Festschrift}, Springer.


% 
% \bibitem{Elk} El Karoui N., Quenez M. (1995): Dynamic programming and
% pricing of contingent claims in an incomplete market, {\it SIAM J. Control and
% Optimization}, {\bf33}, 29-66.

\bibitem{Follmer}F\"{o}llmer H., Leukert P. (1999): Quantile hedging,
{\it Finance and Stochastic}, {\bf3}, 251-273.

\bibitem{Fouque}Fouque J. P., Papanicolaou G., Sircar K. R. (2000): {\it Derivatives in Financial Markets with Stochastic Volatility}, Cambridge University Press.

\bibitem{Fried}Friedman A. (1975): {\it Stochastic differential
equations and applications}, vol.1, Academic Press.

\bibitem{Gam}Gamys M., Kabanov Y. (2009): Mean square error for the
Leland--Lott hedging strategy, {\it Recent Advances in Financial Engineering}, Proceedings of the 2008 Daiwa Int. Worshop on Financial Engineering, World Scientific.

\bibitem{Grand} Granditz P., Schachinger W. (2001): Leland's approach to
option pricing: The evolution of discontinuity, {\it Mathematical Finance}, {\bf 11}, 347-355.

% \bibitem{Gobet} Gobet E., Temam E. (2001): Discrete time hedging errors for option with irregular payoffs, {\it Finance and Stochastics}, {\bf5}, 357-367.

\bibitem{Hall}Hall P. (1980): {\it Martingale limit theory and its
applications}, Academic Press.

\bibitem{Heston}Heston S. (1993): A closed-form solution for options with stochastic volatility with applications to bond and currency options, {\it Rev. Fin. Stud.}, {\bf6}(2), 327-343.

\bibitem{IbHa}Ibragimov A. I., Hasminskii Z. R. (1981): {\it Statistical estimation: asymptotic theory}, English transl. by Samuel Kotz,
 Springer Verlag-Berlin.

\bibitem{Kab-Saf1} Kabanov Y., Safarian M. (1997): On Leland's strategy
of option pricing with transaction costs, {\it Finance and Stochastics}, {\bf1}
, 239-250.

\bibitem{Kab-Saf2} Kabanov Y., Safarian M. (2009): {\it Markets with
transaction costs: Mathematical Theory}, Springer - Verlag Berlin.

\bibitem{Kab-Ser} Kabanov Y., Pergamenshchikov S. (2003): {\it Two-scale stochastic systems, Asymptotic Analysis and Control}, Springer - Verlag Berlin.

\bibitem{Kallsen} Kallsen J., Muhle-Karbe J. The general structure of optimal investment and consumption with small transaction costs, {\it preprint}.


\bibitem{Kar}Karatzas I., Shreve S. (1998): {\it Methods of mathematical
finance}, Springer-Verlag Berlin.

%\bibitem{Kutoyants} Yu.A. Kutoyants (2004). \textit{Statistical inference
%for ergodic diffusion processes}. Springer - Verglag London.

\bibitem{Leland} Leland H. (1985): Option pricing and replication with
transactions costs, {\it Journal of Finance}, {\bf40}, 1283-1301.

\bibitem{Lepinette08}L\'epinette E. (2008): March\'{e} avec c\^{o}uts de transaction:
approximation de Leland et arbitrage, {\it Th\`{e}se doctorale}, Universit\'{e} de
Franche-Comt\'{e} Besan\c{c}on.

\bibitem{Lepinette09}L\'epinette E. (2009): Leland's approximations for concave
pay-off functions, {\it Recent Advances in Financial Engineering}, World
Scientific.

\bibitem{Lepinette-Kab}L\'epinette E., Kabanov Y. (2010): Mean square error for the
Leland-Lott hedging strategy: convex payoffs, {\it Finance and Stochastics}, {\bf14}(4), 625-667.

\bibitem{Lepinette10}L\'epinette E. (2012): Modified Leland's strategy for constant
transaction costs rate, {\it Mathematical Finance}, {\bf22}(4), 741-752.

\bibitem{Lions-Mus} Lions P.-L., Musiela M. (2007): Correlations and bounds for stochastic volatility models, {\it Ann. I. H. Poincar\'e}, {\bf24}, 1-16.



\bibitem{Lip-Shir}Liptser R. S., Shiryaev N. A. (2001):  {\it Statistics of
random processes I: General Theory},  Applications of Mathematics, Springer -
Verlag Berlin .

\bibitem{Lott} Lott K. (1993): Ein verfahren zur eplikation von optionen
unter transaktionkosten in stetiger Zeit, {\it Dissertation}, Universit\"{a}t der
Bundeswehr M\"{u}nchen.
\bibitem{Nguyen1} Nguyen H. T.:  Option replication with general transaction costs in stochastic volatility markets, {\it preprint}.

\bibitem{Nguyen2} Nguyen H. T.:  Approximate hedging with proportional transaction costs for multi-asset options, {\it submitted}.

\bibitem{Novikov} Novikov A. (1997): Hedging of options with a given
probability, {\it Theory Probability Applications}, {\bf 43}(1), 135--143.

\bibitem{Per}Pergamenshchikov S. (2003): Limit theorem for Leland's
strategy, {\it Annals of Applied Probability}, {\bf 13}, 1099-1118.

\bibitem{Pham} Pham H. (2002): Smooth solutions to optimal investment models with stochastic volatilities and portfolio constraints, {\it Appl. Math. Optim.}, {\bf 46}, 55-78.

\bibitem{PhamTouzi} Pham H., Touzi N. (1996): Equilibrium state prices in a stochastic volatility model, {\it Mathematical Finance}, {\bf 6}, 215-236.

\bibitem{RenaultTouzi} Renault E., Touzi N. (1996): Option hedging and implicit volatilities, {\it Mathematical Finance}, {\bf 6}, 279-302. 
\end{thebibliography}
\end{document}